\documentclass[twocolumn,secnumarabic,amssymb, nobibnotes, aps, pre,superscriptaddress]{revtex4-2}

\usepackage{amsmath}
\usepackage{amsfonts}
\usepackage{amssymb}
\usepackage{graphicx}
\usepackage{hyperref}
\usepackage{longtable}
\usepackage{dcolumn}
\usepackage{bm}
\usepackage{epstopdf}
\usepackage{multirow}
\usepackage{tablefootnote}
\usepackage{cancel}
\allowdisplaybreaks

\usepackage{soul}

\graphicspath{{../figures/}}
\usepackage{color}

\newcommand\bs{\boldsymbol}
\begin{document}

\title{Advecting Scaffolds: Controlling The Remodelling Of Actomyosin With Anillin
}

\author{Denni Currin-Ross}
\affiliation{Centre for Cell Biology of Chronic Disease, Institute for Molecular Bioscience, The University of Queensland, Brisbane 4000, Australia.}
\affiliation{School of Physics, UNSW, Sydney 2052, Australia.}
\affiliation{EMBL Australia Node in Single Molecule Science, School of Biomedical Sciences, UNSW, Sydney 2052, Australia.}

\author{Sami C.~Al-Izzi}
\email{samiali@uio.no}
\affiliation{Department of Mathematics, Faculty of Mathematics and Natural Sciences, University of Oslo, 0315 Oslo, Norway.}

\author{Ivar Noordstra}
\affiliation{Centre for Cell Biology of Chronic Disease, Institute for Molecular Bioscience, The University of Queensland, Brisbane 4000, Australia.}

\author{Alpha S.~Yap}
\affiliation{Centre for Cell Biology of Chronic Disease, Institute for Molecular Bioscience, The University of Queensland, Brisbane 4000, Australia.}

\author{Richard G.~Morris}
\email{r.g.morris@unsw.edu.au}
\affiliation{School of Physics, UNSW, Sydney 2052, Australia.}
\affiliation{EMBL Australia Node in Single Molecule Science, School of Biomedical Sciences, UNSW, Sydney 2052, Australia.}

\date{\today}

\begin{abstract}
    We propose and analyse an active hydrodynamic theory that characterises the effects of the scaffold protein anillin.    
    Anillin is found at major sites of cortical activity, such as adherens junctions and the cytokinetic furrow, where the canonical regulator of actomyosin remodelling is the small GTPase, RhoA.
    RhoA acts via intermediary `effectors' to increase both the rates of activation of myosin motors and the polymerisation of actin filaments. 
    Anillin has been shown to \emph{scaffold} this action of RhoA--- improving critical rates in the signalling pathway without altering the essential biochemistry--- but its contribution to the wider spatio-temporal organisation of the cortical cytoskeleton remains poorly understood.
    Here, we combine analytics and numerics to show how anillin can non-trivially regulate the cytoskeleton at hydrodynamic scales.
    At short times, anillin can amplify or dampen existing contractile instabilities, as well as alter the parameter ranges over which they occur.
    At long times, it can change both the size and speed of steady-state travelling pulses.
    The primary mechanism that underpins these behaviours is established to be the {\it advection} of anillin by myosin II motors, with the specifics relying on the values of two coupling parameters. These codify anillin's effect on local signalling kinetics and can be traced back to its interaction with the acidic phospholipid phosphatidylinositol 4,5-bisphosphate (PIP$_2$), thereby establishing a putative connection between actomyosin remodelling and membrane composition.     
\end{abstract}

\maketitle

\section{Introduction}\label{sec:intro}

The actomyosin cortex is a cytoskeletal component and a canonical example of an active adaptive composite material \cite{Prost2015ActivePhysics,Banerjee2020TheMaterial}. It comprises a complex architecture of actin filaments, myosin-II motors and over a hundred actin-binding proteins (ABPs) \cite{Chugh2018TheGlance}. Due to the regulated synthesis and degradation of these components, as well as the conversion of chemical energy into mechanical forces by molecular motors, actomyosin is able to dynamically remodel, changing its shape as well as material properties \cite{Prost2015ActivePhysics,Banerjee2020TheMaterial}. It therefore plays a critical role in many mechanical aspects of cellular and tissue function, including cell migration \cite{vedula2014epithelial}, division \cite{piekny2005cytokinesis,maitre2016asymmetric} and adhesion junction formation \& maintenance \cite{ratheesh2012centralspindlin,arnold2017rho}.

\begin{figure*}[t]
    \centering
    \includegraphics[width=\textwidth]{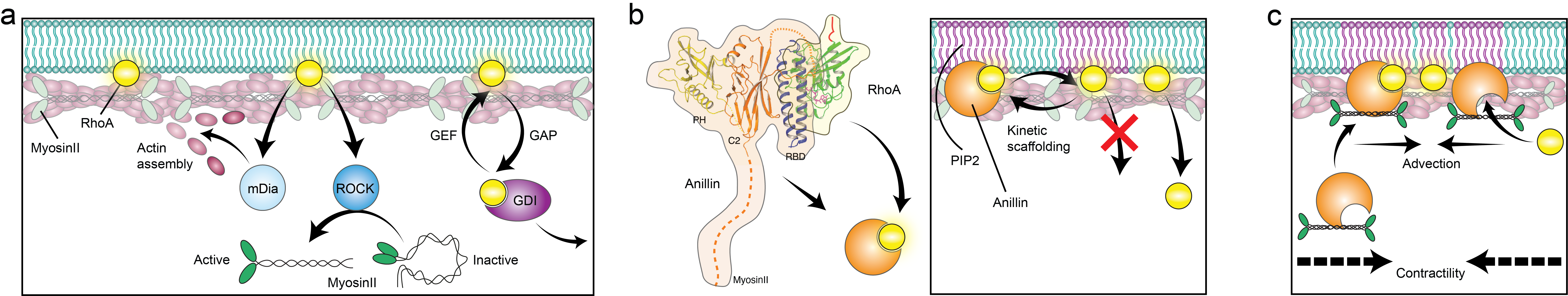}
    \caption{
        \label{fig:RhoASignallingPathway}
        \textbf{Understanding RhoA signalling and the action of the scaffold, anillin.} \textbf{a,} Canonical RhoA signalling pathway. RhoA cycles between an active Guanine Triphosphate (GTP) bound state and an inactive Guanine Diphosphate (GDP) bound states, mediated by Guanine Exchange Factors (GEFs) and GTPase Activating Proteins (GAPs), respectively. Inactive RhoA is extracted and sequestrated from the membrane via cytoplasmic GDP Dissociation Inhibitors (GDIs). Membrane-active RhoA binds downstream effectors mDia and ROCK to promote actin assembly and myosin activation in order to modulate cortical activity. \textbf{b,} Protein structure of anillin-RhoA binding, modified from PDB ID 1s1c \cite{dvorsky2004always}. Anillin contains N-terminal F-actin and non-muscle myosin II binding domains as well as anillin homology (AH: C2-RBD) and pleckstrin homology (PH) domains at the C terminus. RhoA and effectors mDia1 and ROCK1 bind to the same AH domain at the C-terminus. Since anillin cannot simultaneously bind RhoA and its effectors, it cannot act as a traditional tether scaffold. Instead, transient binding of active RhoA to anillin promotes its entry into a complex with phosphoinositide-4,5-P2 (PIP$_2$) which, when unbound, counters its dissociation from the membrane whilst remaining able to bind its downstream effectors, mDia1 and ROCK1. Cycles of such transient binding therefore enhance the residence time and ultimately signalling output of active RhoA. \textbf{c,} Myosin-anillin anchorage. Anillin can bind to active non-muscle myosin II motors \cite{straight2005anillin}. Assuming that myosin-bound anillin is advected under contractile remodelling then gives rise to a novel mechanism for spatially modulating RhoA signalling in the actomyosin cortex.
        }
\end{figure*}

The small GTPase RhoA functions as an essential upstream regulator at these critical sites of cortical activity \citep{banerjee2017actomyosin, hall1998rho}. In its active form, GTP-RhoA binds and activates otherwise autoinhibited contractile `effectors', such as the kinase, ROCK1 and the formin, mDia1 (Fig.~\ref{fig:RhoASignallingPathway}a) \cite{jaffe2005rho}.  Since the former increases phosphorylation the myosin regulatory light chain (MRLC), and the latter promotes actin polymerisation, local increases in RhoA result in greater cortical contractility (Fig.~\ref{fig:RhoASignallingPathway}a) \cite{hall1998rho}.

In the traditional biological picture, it is the spatial and temporal localization of these contractile regulators that permits coordinated contractility, and therefore control over cellular morphology. However, the precision with which a cell can reliably control the concentration of RhoA is limited by a number of factors. Most notably, the relatively long timescales of expression, spatio-temporal limits of the inter-cellular trafficking apparatus, a labile association with the membrane \cite{Budnar2019AnillinKinetics}, and rapid diffusion due to a low molecular weight \cite{henis2006frap}.

Such issues were partially addressed in a recent publication \cite{Budnar2019AnillinKinetics}, whereby the protein anillin was shown to \emph{scaffold} the action of RhoA--- {\it i.e.}, anillin improves the kinetics of RhoA's signalling pathways without altering their biochemistry. Anillin does so by repeatedly co-localising cortical RhoA with the acidic membrane phospholipid, phosphatidylinositol 4,5-bisphosphate (PIP$_2$). This has the effect of increasing the cortical dwell time of RhoA without binding to, and hence blocking, the key binding site needed to actively engage its effectors. As such, anillin improves the efficacy of RhoA signalling, and helps overcome the otherwise rapid turnover rate of active RhoA at the membrane (Fig.~\ref{fig:RhoASignallingPathway}b).

However, if the anillin-PIP$_2$ pathway stabilises RhoA-signalling temporally, then this begs the question: what localises it, spatially? Moreover, how do such mechanisms couple to the active-contractile-advective feedback that is characteristic of actomyosin?

In two pioneering works, differing applications of non-linear reaction-diffusion dynamics have been proposed to explain the interplay between RhoA and pattern formation of actomyosin \cite{nishikawa2017controlling,staddon2022pulsatile}.  In \cite{nishikawa2017controlling} a phenomenological approach was used, invoking the Swift-Hohenberg oscillator \cite{cross1993pattern} to describe RhoA dynamics. Reference \cite{staddon2022pulsatile}, by contrast, identifies specific mechanisms: notably, the autocatalysis of RhoA activation and the F-actin dependence of GTPase Activating Proteins (GAPs) \cite{michaux2018excitable}.

Here, we describe how anillin can provide a third, scaffolding-based alternative. Motivated by \cite{Budnar2019AnillinKinetics}, the structure of anillin (PDB ID 1s1c \cite{dvorsky2004always}), and its known binding to myosin II \cite{straight2005anillin}, we propose a long wavelength, hydrodynamic model of the cortex that explicitly couples actomyosin mechanics to the RhoA-effector signalling pathway via anillin advection (Fig.~\ref{fig:RhoASignallingPathway}c).

Our results are organised as follows. In Section \ref{sec:actomyosin}, we outline a  scalar active hydrodynamic theory of actomyosin. In this, the (first order) kinetics of myosin activation and actin polymerisation arise due to the concentrations of mDia1 and ROCK1, respectively, which are assumed to be downstream of a constant RhoA signal. We identify the relevant dimensionless parameters, the conditions for stability/instability in the linear regime, and the exceptional points of the system. Using parameter values taken from the literature, finite-difference numerics are used to capture behaviours beyond the linear regime. The system is initially excitatory-inhibitory in nature, despite linear kinetics, because of an instantaneous Stokesian force balance condition. At long times, however, our model settles to form steady-state travelling pulses. In Section \ref{sec:level2}, we incorporate the effects of the anillin-RhoA pathway \cite{Budnar2019AnillinKinetics} via the anchoring of anillin to myosin II motors. This allows us to contrast the extended system involving anillin scaffolding with that of canonical RhoA dynamics. On short times, the advection of anillin due to its coupling to myosin II can significantly alter the system's hydrodynamic instabilities, amplifying or dampening them, as well as changing the parameter values for which they occur. On long times, numerical simulations show that the RhoA-anillin pathway increases the size and speed of steady state travelling pulses.  The details rely on the values of two coupling constants, whose potential significance is treated in the Discussion, since they have previously been identified with the acidic phospholipid phosphatidylinositol 4,5-bisphosphate (PIP$_2$), thereby establishing a potential connection between actomyosin remodelling and membrane composition.

\section{Scalar Active hydrodynamics of actomyosin}\label{sec:actomyosin}

Neglecting the polar nature of the actin meshwork, we characterise the underlying physics of isotropic actomyosin by only two scalar fields: the density of actin, $\rho_{\mathrm{a}}(x,t)$, and the density of myosin II molecules bound to the actin meshwork, $\rho_{\mathrm{m}}(x,t)$. This precludes the formation of ordered patterns, such as the asters, spirals and vortices reported elsewhere \cite{kruse2004asters, kruse2005generic}, or the notion of directed procession of myosin motors along filaments \cite{mosby2020myosin,al2021more}, and instead permits us to focus on the impact of the RhoA signalling pathway and the action of anillin. In one spatial dimension, the dynamics of our model is given by two continuity equations:
\begin{equation}\label{eq:rhoaC}
    \partial_t\rho_{\mathrm{a}} + \partial_x \left(v\rho_{\mathrm{a}}\right) = S_a\textrm{,}
\end{equation} 
and
\begin{equation}\label{eq:rhomC}
    \partial_t\rho_{\mathrm{m}} + \partial_x \left(v\rho_{\mathrm{m}}\right) - D_m\partial_x^2\rho_{\mathrm{m}} = S_m\textrm{,}
\end{equation}
where $S_a$ and $S_m$ are source and sink terms associated with actin assembly/disassembly and myosin binding/unbinding, respectively. Here, $v$ represents the local velocity of the actin meshwork, and $D_m\partial_x^2\rho_{\mathrm{m}}$ is a diffusive flux of myosin, which is assumed to capture the phenomenon of myosin motors `walking' along (isotropically ordered) actin filaments \cite{mosby2020myosin}.

Equations~(\ref{eq:rhoaC}) and (\ref{eq:rhomC}) are not closed, and must be augmented by a force-balance equation which specifies the actin velocity, $v$. We assume a Rouse-like limit, where the divergence of the total stress, $\sigma$, is balanced by a non-momentum conserving friction
\begin{equation}\label{eq:force}
     \Gamma v = \partial_x\sigma\textrm{,}
\end{equation}
with $\Gamma$ a cortical friction coefficient. We write $\sigma$ as a sum of active and passive parts
\begin{equation}\label{eq:sigma}
    \sigma = \eta \partial_x v -\chi_a \rho_{\mathrm{a}} + \frac{\zeta \rho_{\mathrm{m}}}{\rho_{\mathrm{m}} + \rho_{\mathrm{m}}^*}\textrm{,}  
\end{equation}
where $\eta$ is the cortex viscosity, $\chi_a$ is the inverse compressibility of the actin network \cite{kruse2006contractility} and $\zeta$ is the active contractile strength (which scales with the chemical potential associated with ATP hydrolysis \cite{koster2016actomyosin}). The non-linear dependence of the final term ensures that the active stress saturates, and the passive resistance to compression becomes dominant, at high densities of myosin II (\textit{i.e.}, $\rho_{\mathrm{m}}/\rho_{\mathrm{m}}^\ast \gg 1$).

Constant background RhoA signalling can be implicitly incorporated into the system via basal rate constants for the density of downstream effectors ROCK1 and mDia1. This assumes an abundance of effectors in the cytosol, as well as transient enzymatic activity such that they are not materially advected by cortical remodelling. The resulting separation of time-scales leads us to write the rate of myosin activation via phosphorylation as $k_{\mathrm{on}}\rho_{\mathrm{ROCK}}$ and the rate of filamentous actin assembly via polymerisation as $k_{\mathrm{p}}\rho_{\mathrm{mDia}}$ (Fig.~\ref{fig:RhoASignallingPathway}a). Once phosphorylated, myosin can bind at a rate dependent on the density of available sites on the actin network, $\rho_{\mathrm{a}} - \rho_{\mathrm{m}}$. Similarly, myosin inactivates and actin disassembles at the constant, per unit mass, rates of $k_{\mathrm{off}}$ and $k_{\mathrm{dp}}$. The resultant mass action kinetics are then captured by
\begin{equation}\label{eq:source}
    S_m = k_{\mathrm{on}}\rho_{\mathrm{ROCK}}\left(\rho_{\mathrm{a}} - \rho_{\mathrm{m}}\right) - k_{\mathrm{off}}\rho_{\mathrm{m}}\textrm{,}
\end{equation}
and
\begin{equation}\label{eq:sink}
    S_a = k_{\mathrm{p}}\rho_{\mathrm{mDia}} - k_{\mathrm{dp}}\rho_{\mathrm{a}}\textrm{,}
\end{equation}
where, to ensure a biologically valid setting, we impose that actin disassembly cannot exceed the myosin inactivation rate--- \textit{i.e., }$k_{\mathrm{off}} + k_{\mathrm{on}} - k_{\mathrm{dp}} > 0$.

\begin{figure*}[th!]
    \includegraphics[width=\textwidth]{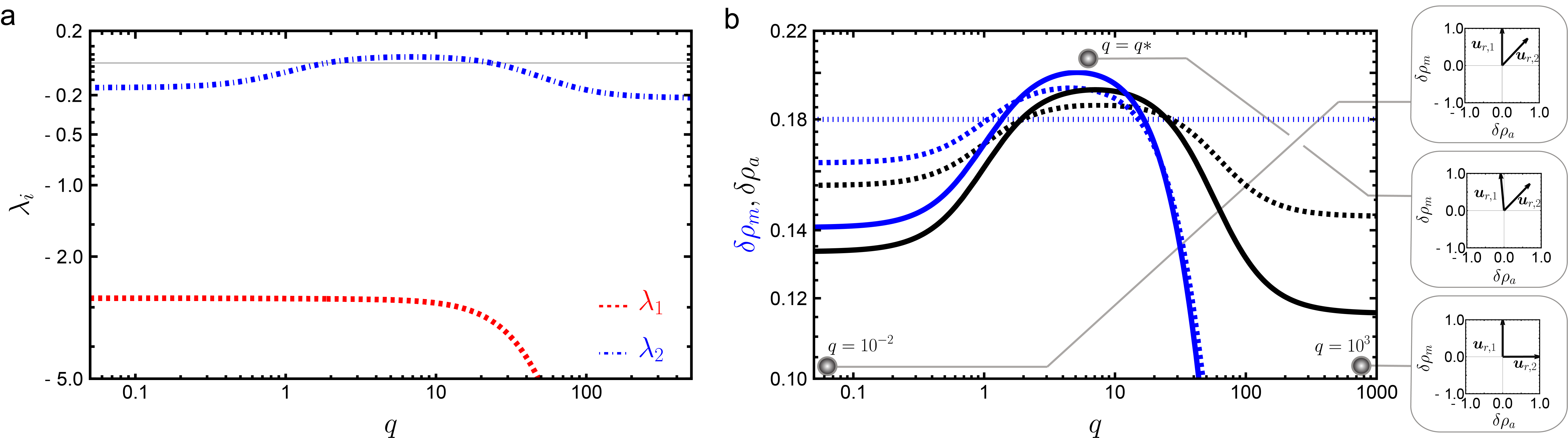}
    \caption{
        \label{fig:linearResponseAnalysis}
        \textbf{Eigensystem of scalar actomyosin}. \textbf{a,} Real eigenvalues, $\lambda_{1,2}$ of the linear response matrix $\mathsf{A}$ [Eq.~(\ref{eq:linearResponseMatrix2D})].
        For representative parameter values taken from the literature (Table.~\ref{tab:ParameterValues}) only the second eigenvalue, $\lambda_2$, is positive, corresponding to a `contractile-kinetic' instability [{\it i.e.}, Eq.~(\ref{eq:eigen2})]. \textbf{b,} Forward time-evolution of the linear response matrix $\mathsf{A}$ [Eq .~(\ref{eq:linearResponseMatrix2D})], plotted for parameters in Table~\ref{tab:ParameterValues}. Dotted, dashed and solid lines denote (dimensionless) time points $t = 0$, $t = 1$, and $t = 2$, respectively. Inserts show the right-eigenvectors, $\mathbf{u}_r$, plotted for $q=10^{-2}$ and $q=10^3$ ({\it i.e.}, large and small wavelengths, respectively) as well as at the wavenumber of maximum growth, $q^*\approx 7.0$.  At wavenumbers both smaller and of similar magnitude to $q^\ast$, the positivity of $\lambda_2$ [Eq.~(\ref{eq:eigen2})] drives changes in {\emph both} $\delta\rho_{\mathrm{m}}$ and $\delta\rho_{\mathrm{a}}$ (\textit{i.e.}, $\boldsymbol{u}_{r,2}$ forms an angle of approximately $\pi/2$ with the positive x-axis). At larger wavenumbers, however, $\delta\rho_{\mathrm{m}}$ and $\delta\rho_{\mathrm{a}}$, decouple, with their $q$-dependent growth rate (or absence thereof) being dictated by the positivity of $\lambda_1$ and $\lambda_2$, respectively [\textit{i.e.}, conditions (\ref{eq:eigen1}) and (\ref{eq:eigen2})].  
    }
\end{figure*}

\subsection{Linear Stability}
%
The homogenous steady state solution to our model is given by $\rho^0_\mathrm{a}=k_{\mathrm{p}}\rho_{\mathrm{mDia}}/k_{\mathrm{dp}}$ and $\rho^0_{\mathrm{m}} = k_{\mathrm{on}}\rho_{\mathrm{ROCK}}\rho^0_{\mathrm{a}}/(k_{\mathrm{on}}\rho_{\mathrm{ROCK}}+k_{\mathrm{off}})$, which ensures that $\rho^0_\mathrm{a}\ge\rho^0_\mathrm{m}$. To understand the stability around this state, we linearize in terms of small perturbations of the form $\rho_{\mathrm{m}}(x,t) = \rho_{\mathrm{m}}^0 + \delta\rho_{\mathrm{m}}(x,t)$, $\rho_{\mathrm{a}}(x,t) = \rho_{\mathrm{a}}^0 + \delta\rho_{\mathrm{a}}(x,t)$ and $v(x,t) = \delta v(x,t)$. Re-scaling the spatial and temporal coordinates by $x \to x\sqrt{\eta / \Gamma}$ and $t \to t$ $\eta/\zeta$, where $\sqrt{\eta / \Gamma}$ represents the hydrodynamic length of the cortex, our dimensionless linear equations are then
\begin{equation}
\begin{aligned}
  &\partial_t\delta\rho_{\mathrm{m}} + \partial_x\delta v - \text{Pe}^{-1}\partial_x^2\delta\rho_{\mathrm{m}} = \Hat{k}_{\mathrm{on}}\delta\rho_{\mathrm{a}} - \Tilde{k}_{\mathrm{off}}\delta\rho_{\mathrm{m}}\textrm{,} \\
  &\partial_t\delta\rho_{\mathrm{a}} + \partial_x \delta v = - \Tilde{k}_{\mathrm{dp}}\delta\rho_{\mathrm{a}}\textrm{,} \\
  & \partial_x^2\delta v - \delta v - \Tilde{\chi}_a\partial_x\delta\rho_{\mathrm{a}} + h(\rho_{\mathrm{m}}^*) \partial_x \delta\rho_{\mathrm{m}} = 0\textrm{,}
\end{aligned}
\end{equation}
where re-scaled quantities are defined in Table~\ref{tab:AMParameters}. Analysing the system in terms of the wavenumber $q$, defined by the Fourier transform $F(x)=\int\mathrm{d}q/(2\pi)\bar{F}(q)\exp(i q x)$, we find the following dynamical equations in reciprocal space
\begin{equation}
  \partial_t{\bs{\delta \Bar{\rho}}}_{q} = \mathsf{A}\cdot{\bs{\delta \Bar{\rho}}}_{q}\textrm{,}\label{eq:reciprocal}
\end{equation}
where ${\bs{\delta \Bar{\rho}}}_q= (\delta\Bar{\rho}_{a},\delta\Bar{\rho}_{m})^\mathrm{T}$ and
\begin{equation}\label{eq:linearResponseMatrix2D}
   \mathsf{A} =  \left( \begin{matrix}
   -\Tilde{k}_{\mathrm{dp}} -\Tilde{\chi}_a\frac{q^2}{q^2+1} &\frac{q^2 h(\rho_{\mathrm{m}}^*)}{q^2+1}\\
        \Hat{k}_{\mathrm{on}} -\Tilde{\chi}_a\frac{q^2}{q^2+1} &  - \Tilde{k}_{\mathrm{off}} -\Tilde{D}_m  q^2+\frac{q^2 h(\rho_{\mathrm{m}}^*)}{q^2+1}
    \end{matrix}\right).
\end{equation}
%
\begin{table}[t]
    \caption{\label{tab:AMParameters}
    Non-dimensionalised parameters used in both the linear stability and non-linear numerical analysis of the isotropic model of actomyosin.}
    \begin{ruledtabular}
        \begin{tabular}{cc}
            Non-dimensionalised Parameter &  Rescaled quantity\\
            \hline
            $\Tilde{k}_{\mathrm{off}}$& $(k_{\mathrm{off}} + k_{\mathrm{on}}\rho_{\text{ROCK}})\eta / \zeta$\\
            $\Tilde{k}_{\mathrm{p}}$& $k_{\mathrm{p}}\rho_{\text{mDia}}\eta / \zeta\rho_{\mathrm{a}}^0$\\
            $\Tilde{k}_{\mathrm{dp}}$& $k_{\mathrm{dp}}\eta / \zeta$\\
            $\Hat{k}_{\mathrm{on}}$& $k_{\mathrm{on}}\rho_{\text{ROCK}}\eta\rho_{\mathrm{a}}^0 / \zeta\rho_{\mathrm{m}}^0$\\
            \text{Pe} & $\zeta/D_m\Gamma$\\
            $\Tilde{\chi}_a$& $\chi_a\rho_{\mathrm{a}}^0 / \zeta$\\
            $h(\rho_{\mathrm{m}}^*)$& $\rho_{\mathrm{m}}^0\rho_{\mathrm{m}}^* / (\rho_{\mathrm{m}}^0+\rho_{\mathrm{m}}^*)^2$\\
        \end{tabular}
    \end{ruledtabular}
\end{table}
%
The solution to (\ref{eq:reciprocal}) is of the form
\begin{equation}
    {\bs {\delta\Bar{\rho}}}_q(t)= e^{\mathsf{A}t}\cdot{\bs {\delta \Bar{\rho}}}(0) = c_1 e^{\lambda_1 t}{\bs u}_1 + c_2 e^{\lambda_2 t}{\bs u}_2,
\end{equation}
where ${\bs {\delta\Bar{\rho}}}(0)=c_1 {\bs u}_1 + c_2 {\bs u}_2$ is the initial Fourier data of the system, $\lambda _{1,2}$ are the eigenvalues of $\mathsf{A}$ and ${\bs u}_{1,2}$ are the corresponding right eigenvectors. The eigenvalues can be calculated, with full length expressions provided in Eq.~(\ref{eq:eigenvalues_FL}) of Appendix~\ref{sec:SuppLinearStability2D}, and plotted in panel {\bf a} of Fig.~\ref{fig:linearResponseAnalysis}. Retaining only terms up to $\mathcal{O}(q^2)$, we find that the conditions for $\lambda_1$ and $\lambda_2$ to be positive, and hence for the system to be unstable for small wavenumbers ({\it i.e.}, long wavelengths), are respectively given by
\begin{equation}\label{eq:eigen1}
        \frac{\rho_{\mathrm{m}}^0\rho_{\mathrm{m}}^*}{(\rho_{\mathrm{m}}^0+\rho_{\mathrm{m}}^*)^2} \left(1 - \frac{\Hat{k}_{\mathrm{on}}}{\Tilde{k}_{\mathrm{off}} - \Tilde{k}_{\mathrm{dp}}} \right) - \mathrm{Pe}^{-1} > 0 \textrm{ ,}
\end{equation}
and
\begin{equation}\label{eq:eigen2}
        -\Tilde{\chi}_a + \frac{\Hat{k}_{\mathrm{on}{}}}{\Tilde{k}_{\mathrm{off}} - \Tilde{k}_{\mathrm{dp}}}  > 0\textrm{ ,}
\end{equation}
where $\mathrm{Pe}=$ $\zeta/D_m\Gamma$ is the P\'eclet number. This characterises the relative magnitude of the advection of myosin motors due to active contractility as compared to their motion due to diffusion, where the latter arises athermally in our model due to the capacity of myosin II motors to process, or `walk' along actin filaments.

Importantly, by taking representative parameter values from the literature (Table ~\ref{tab:AMParameters}) we see that only condition (\ref{eq:eigen2}) is unstable, with condition (\ref{eq:eigen1}) remaining stable for all $q$ (Fig.~\ref{fig:linearResponseAnalysis}a).

\subsection{Non-Hermiticity}

The interpretation of (\ref{eq:eigen2}) is complicated by the fact that the response matrix $\mathsf{A}$ is non-Hermitian, which means that right and left eigenvectors are not equivalent. Moreover, the right eigenvectors $\boldsymbol{u}_{r,1}$ and $\boldsymbol{u}_{r,2}$, to which the stability conditions (\ref{eq:eigen1}) and (\ref{eq:eigen2}) pertain, respectively, are not orthogonal in the phase-space spanned by $\delta\bar{\rho}_{\mathrm{m}}$ and $\delta\bar{\rho}_{\mathrm{a}}$. In our case, the difference between the directions of $\boldsymbol{u}_{r,1}$ and $\boldsymbol{u}_{r,2}$ also depends on $q$.

We therefore apply the forward time-evolution operator, $\exp(\mathsf{A}t)$, in order to understand the impact of the instability (\ref{eq:eigen2}) on both $\delta\bar{\rho}_{\mathrm{a}}$ and $\delta\bar{\rho}_{\mathrm{m}}$ (Fig.~\ref{fig:linearResponseAnalysis}b). This shows that, above a critical lengthscale, corresponding to $q\lesssim 10$, the growth of both actin and myosin densities have the same $q$-dependence ({\it i.e.}, they are slaved to one-another). However, beneath that lengthscale, a perturbation of a given $q$ grows differently depending on whether it was to the density of actin or myosin. The density of myosin ultimately becomes stable--- {\it i.e.}, it does not grow--- when subject to high-$q$ perturbations.

\begin{figure*}[t!]
    \centering
    \includegraphics[width=\textwidth]{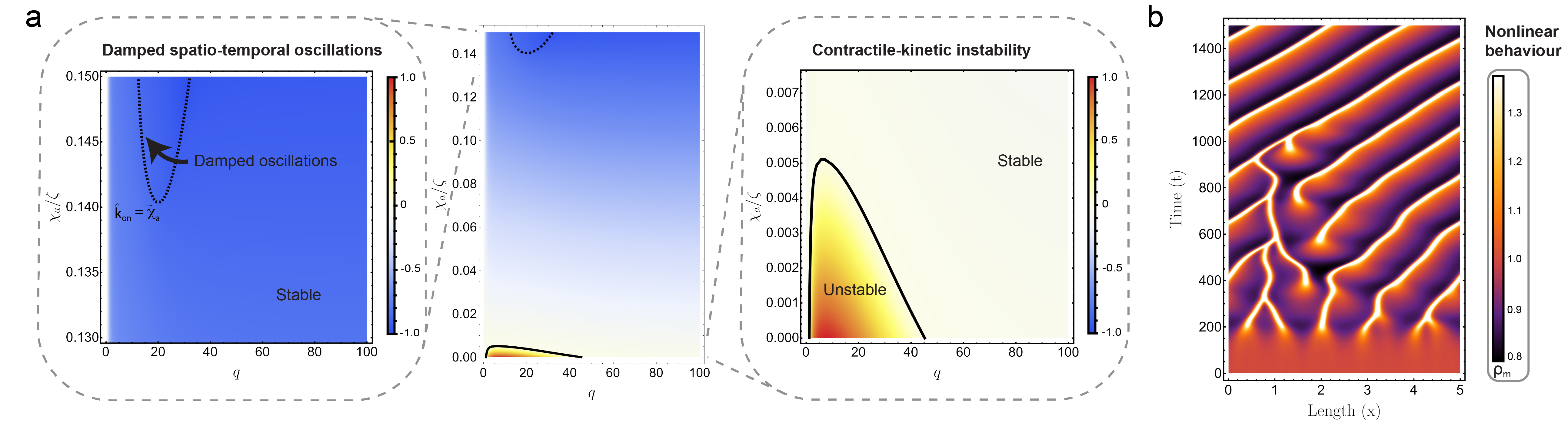}
    \caption{\textbf{Scalar actomyosin; contractile-kinetic instability, damped oscillations, and steady state pulses.} \textbf{a,} Linear stability diagram showing the ratio of inverse compressibility to active contractility, $\chi_a / \zeta$, as a function of wavenumber, $q$, plotted for representative values of all other parameters (Table ~\ref{tab:ParameterValues}). The `contractile-kinetic' instability (\ref{eq:eigen2}) corresponds to small values of $\chi_a / \zeta$ (magnified, right). Whereas, damped oscillations (\ref{eq:highQ_EP_condtion}) correspond to large values of $\chi_a / \zeta$ (magnified, left). \textbf{b,} Kymograph of the numerical solution of our system within the unstable contractile-kinetic regime. From randomised initial conditions, a transient and excitable regime demonstrates foci nucleation and merger before eventually giving way to a three-peak travelling front.}
    \label{fig:Actomyosin_dynamics}
\end{figure*}

This intuition is confirmed by analysing $\boldsymbol{u}_{r,1}$ and $\boldsymbol{u}_{r,2}$ directly. Relegating the full forms to Appendix~\ref{sec:SuppLinearStability2D}, expansions for both small and large $q$, respectively, give the following results: 
\begin{widetext}
\begin{align} \label{eq:spatialLimitsLongWavelength}
     q \ll 1: \quad &   {\bf u}_{r,1} = \left(\text{ } 0\text{ }, \text{ } 1 \text { }\right)^\mathrm{T}   \ & {\bf u}_{r,2} &= \left( \frac{\Tilde{k}_{\mathrm{off}} - \Tilde{k}_{\mathrm{dp}} }{\Hat{k}_{\mathrm{on}}\sqrt{1 + \left(\frac{\Tilde{k}_{\mathrm{off}} - \Tilde{k}_{\mathrm{dp}}}{\Hat{k}_{\mathrm{on}}}\right)^2}}, \frac{1}{\sqrt{1 + \left(\frac{\Tilde{k}_{\mathrm{off}} - \Tilde{k}_{\mathrm{dp}}}{\Hat{k}_{\mathrm{on}}}\right)^2}} \right)^\mathrm{T}\\\label{eq:spatialLimitsShortWavelength}
     q \gg 1: \quad &  {\bf u}_{r,1} = \left(\text{ } 0\text{ }, \text{ } 1 \text { }\right)^\mathrm{T}   \ & {\bf u}_{r,2} &= \left(\text{ } 1 \text{ }, \text{ } 0 \text{ } \right)^\mathrm{T}
\end{align}
\end{widetext}
where it is immediately clear that the eignevectors are not orthogonal in the low $q$ regime, and that their relative directions changes with $q$. In between the limits of small and large $q$, we can also identify the mode which grows the fastest
\begin{equation}\label{eq:qmax2D}
    q^* = \arg\max_{q}\{\lambda_{1,2}\}.
\end{equation}
although the analytical form of both $q^\ast$ and the corresponding eigenvectors were sufficiently unwieldly to calculate.

The directions of the eignenvectors in these three regimes--- small $q$, large $q$, and maximum growth $q$--- can nevertheless been visualised by using the representative parameter values in Table~\ref{tab:AMParameters} (Fig.~\ref{fig:linearResponseAnalysis}c). For $q\ll1$, we have used a value of $q=10^{-2}$, and for $q\gg 1$, we have used $q=10^3$. The value of $q^*$ can be calculated numerically, which results in a value of $\approx7.0$, or a wavelength of $L\approx9.0 \ \mu\text{m}$.
As expected, we see that eigenvector $\boldsymbol{u}_{r,2}$, which corresponds to the physically relevant instability condition (\ref{eq:eigen2}), forms an angle of approximately $\pi/2$ with the positive x-axis for both $q^\ast$ and $q=10^{-2}$.  Since the relevant stability criterion (\ref{eq:eigen2}) concerns the small $q$ regime, we conclude that, when unstable, perturbations to both the myosin and actin densities grow at approximately equal rates. For larger values of $q$--- \textit{i.e.}, small wavelengths--- we see that $\boldsymbol{u}_{r,2}$ drives only the growth of $\delta\bar{\rho}_{\mathrm{a}}$ and is decoupled from that of $\delta\bar{\rho}_{\mathrm{m}}$.  

In this context, the instability that underpins (\ref{eq:eigen2}) can be interpreted as `contractile-kinetic' in the following sense. If the rate by which available sites on actin filaments bind to myosin II head-groups is matched by the rate at which they become available (which is just the rate by which myosin head-groups unbind minus the rate of actin depolymerisation) then the second, kinetic term in (\ref{eq:eigen2}) is just equal to 1. As a result, the system is unstable if the active contractiliy, $\xi$, exceeds the elasticlike resistance to compression, which is characterised by the inverse compressibility, $\chi$, multiplied by the constant $\rho_a^0$.  If, however, myosin motors bind with a higher (lower) frequency than free sites become available, then more (less) myosins can bind in response to increases in actin density due to contracility. This results in a kinetic term that is larger (smaller) than 1, and the instability condition can be met with a lower (higher) value of active contractility.

\subsection{Exceptional Points}

There are two exceptional points in our system, where the eigenvectors align and their corresponding eigenvalues become degenerate. These correspond to high and low $q$ regimes, although the  exceptional point in the low $q$ regime is unphysical since it requires breaking the constraint that $k_{\mathrm{off}} + k_{\mathrm{on}} - k_{\mathrm{dp}} > 0$. For this reason it does not appear in panel {\bf a} of Fig.~\ref{fig:linearResponseAnalysis} and  is excluded from further analysis [see Eqs.~(\ref{eq:UnphysicalEP_lengthscale}) \& (\ref{eq:UnphysicalEP_condition}) in the Appendix~\ref{sec:SuppLinearStability2D} for more details].

In the high $q$ regime, the exceptional point corresponds to the appearance of non-zero imaginary components of the remaining eigenvalue. Since the real part of the eigenvalue is negative, this is indicative of stable ({\it i.e.}, damped) spatio-temporal waves.

Performing a series expansion of the eigenvalues around high $q$ and then equating them, we can identify a upper critical wavenumber, $q^\dagger$, at which such waves occur,
\begin{equation}\label{eq:highQ_EP_lengthscale}
    q^\dagger = \sqrt{\mathrm{Pe}\left[-\Tilde{k}_{\mathrm{off}} +\Tilde{k}_{\mathrm{dp}} +h(\rho_{\mathrm{m}}^*) +\Tilde{\chi}_a\right]},
\end{equation}
which holds only when $h(\rho_{\mathrm{m}}^*) + \Tilde{\chi}_a > \Tilde{k}_{\mathrm{off}} - \Tilde{k}_{\mathrm{dp}}$. Substituting (\ref{eq:highQ_EP_lengthscale}) into the previously calculated expressions for the eigenvalues, leads to the following upper criterion for the exceptional point
\begin{equation}\label{eq:highQ_EP_condtion}
    \Tilde{\chi}_a = \Hat{k}_{\mathrm{on}} \text{.}
\end{equation}
That is, when the (dimensionless) ratio of inverse compressibility to contractility coincides with the (dimensionless) myosin binding rate. Since this is the upper criterion, damped spatio-temporal waves can also be found in a small regime where the strength of the dimensionless inverse compressibility exceeds the myosin binding rate.

The full stability diagram can be found in Fig.~\ref{fig:Actomyosin_dynamics}a, which shows the regimes for both the contractile-kinetic instability [{\it cf.} Eq.~(\ref{eq:eigen2})] and damped spatio-temporal waves [{\it cf.} Eq.~(\ref{eq:highQ_EP_lengthscale})]

\subsection{Non-linear behaviour}

In order to examine the full non-linear behaviour of our model, we resort to numerically solving our system of PDEs. We employ a central space, RK2-time finite difference scheme on a periodic domain of size $L$. The actin and myosin densities were initialised with a small perturbation of the form $\delta\rho_{a,m}=A\sin\left(2\pi x/L + \theta\right)$ around the homogeneous steady state, where the coefficients $A$ and $\theta$ were randomly sampled according to $A\sim \mathcal{U}[0,\text{0.01})$ and $\theta\sim \mathcal{U}[0,2\pi)$.

The numerical solution of our model confirms that the growth of the initial perturbations has excellent quantitative agreement with the linear stability analysis over the first $t/\tau<75$ where $\tau=$ $\eta/\zeta$ is the timescale of viscous dissipation associated with contractile remodelling (Appendix Fig.~\ref{fig:2DAlignment}).

Beyond this time, non-linearities become increasingly important, with the system forming dense foci of actomyosin separated by sparse regions. The transient foci are highly dynamic: newly nucleated foci are attracted to, and merge with larger foci (see kymograph in Fig.~\ref{fig:Actomyosin_dynamics}b and Supplementary video S1) returning to a characteristic size following merger (Appendix Fig.~\ref{fig:FociWidth}a). This excitatory-inhibitory-like behaviour is reminiscent of that previously observed using reaction-diffusion models which rely on nonlinear reaction kinetics \cite{tyson1988singular,bement2015activator,staddon2022pulsatile}. Here, we employ only linear kinetic terms.  However, when coupled via velocities that obey Stokesian dynamics (and hence relax instantaneously) this is sufficient to induce an excitable system \cite{banerjee2017actomyosin}. Despite our stability analysis demonstrating that linearly unstable waves are not possible in our system, long-time dynamics establish a steady travelling front that is reminiscent of the reaction-diffusion models mentioned in the introduction \cite{staddon2022pulsatile,nishikawa2017controlling} and consistent with broader expectations of pulsatile dynamics (Fig.~\ref{fig:Actomyosin_dynamics}).

\begin{table}[t]
    \caption{
        \label{tab:ParameterValues}Parameter values.
        }
    \begin{ruledtabular}
        \begin{tabular}{cll}
            \\
            \multicolumn{3}{c}{Actomyosin}\\
            \\
            \hline
            \hline
            Parameter &  Value & Ref\\
            \hline
            $\sqrt{\eta/\Gamma}$& $10.0 \ \mu m$ & \cite{Saha2016DeterminingCortex}\\
            $\Gamma/\zeta$ & $0.1 \ \mu m^{-2} \ s $& \cite{Saha2016DeterminingCortex}\\
            $D_m$& $0.01 \ \mu m^2s^{-1}$& \cite{nishikawa2017controlling}\\
            $k_{\mathrm{off}}$& $0.06 \ s^{-1}$& \cite{Priya2015FeedbackJunctions}\\
            $k_{\mathrm{on}}$& $0.22 \ s^{-1}$& \cite{Priya2015FeedbackJunctions}\\
            $k_{\mathrm{p}}$& $0.03 \ s^{-1}$& \cite{Kovacs2011N-WASPPathway}\\
            $k_{\mathrm{dp}}$& $0.015 \ s^{-1}$& \cite{Kovacs2011N-WASPPathway}\\
            $\chi_a/\zeta$& $0.0035$ \textsuperscript{\ref{AMA:WEG}}$\ \mu m ^{-1}$ & --\\
            $\rho_{\mathrm{ROCK}}/\rho_{\mathrm{mDia}}$& 0.1 \textsuperscript{\ref{AMA:WEG}}& \cite{gabella1984structural}\\
            \hline
            \hline
            \\
            \multicolumn{3}{c}{Actomyosin-anillin anchorage}\\
            \\
            \hline
            \hline
            Parameter &  Value & Ref\\
            \hline
            $k_{\mathrm{off}}^A$& $0.06 \ s^{-1}$& \cite{beaudet2020importin}\\
            $k_{\mathrm{on}}^A$& $0.22 \ s^{-1}$& \cite{beaudet2020importin}\\
            $\nu$ & 0.125 \footnote{\label{AMA:WEG}Assumption} & --\\
            $\mu$ & 0.2 \textsuperscript{\ref{AMA:WEG}}& --\\
        \end{tabular}
    \end{ruledtabular}
\end{table}

\section{\label{sec:level2} Actomyosin-Anillin Coupling}
%
Recent experimental evidence at both adherens junctions and the cytokinetic furrow have shown that the scaffold protein anillin is needed to support actomyosin remodelling at these important sites of cortical activity \cite{Budnar2019AnillinKinetics,straight2005anillin,field1995anillin}. However, despite possessing domains for binding to the lipid bilayer membrane, as well as for binding RhoA and its effectors mDia1 and ROCK1 \cite{piekny2008anillin}, anillin has been shown to operate in manner different to traditional tether scaffolds, which improve the kinetics of their binding partners by bringing them into close proximity with one-another \cite{good2011scaffold}. Since anillin uses the same PH domain for binding RhoA and its effectors, binding between anillin and RhoA cannot happen simultaneously with binding between anillin and either mDia1 or ROCK1, therefore this mode of action is not possible \cite{Budnar2019AnillinKinetics}.  Instead, recent work \cite{Budnar2019AnillinKinetics} suggests that the anillin's atypical C2 domain acts to localise acidic phospholipids, such as PIP$_2$, which then form an association with active RhoA when it is bound to anillin. The result is that, on unbinding from anillin, the transient PIP$_2$ association antagonises the extraction of RhoA from the membrane. Repeated cycles of binding and unbinding between active RhoA and anillin then have the capacity to significantly increase the dwell time of RhoA at the membrane when it is {\it unbound} to anillin, and therefore able to interact with its downstream effectors.

This mode of scaffolding, which draws analogies with kinetic proofreading and stochastic resetting \cite{morris2020anillin}, improves the efficacy of the RhoA signalling pathway in such a way that increases in the concentration of anillin are {\it linearly} related to increases in the concentration of effectors, such as mDia1 \cite{Budnar2019AnillinKinetics}.  As a result, we may incorporate the effects of anillin into the source and sink terms (\ref{eq:source}) and (\ref{eq:sink}), respectively, via the following modifications, linear in the density of anillin, $\rho_{\mathrm{A}}$:
\begin{equation}\label{eq:source_anillin_myo}
    S_m =  k_{\mathrm{on}}\left(\rho_{\mathrm{a}} - \rho_{\mathrm{m}}\right)(\rho_{\mathrm{ROCK}} + \mu\rho_{\mathrm{A}}) - k_{\mathrm{off}}\rho_{\mathrm{m}}
\end{equation}
\begin{equation}\label{eq:source_aniillin_act}
    S_a = k_{\mathrm{p}}(\rho_{\mathrm{mDia}} + \nu\rho_{\mathrm{A}}) - k_{\mathrm{dp}}\rho_{\mathrm{a}}
\end{equation}
Here, the constants $\mu$ and $\nu$ quantify the anillin-dependent uplift to myosin II activation conferred by ROCK1 and the anillin-dependent uplift in actin polymerisation conferred by mDia1, respectively.

As a result of Eqs.~(\ref{eq:source_anillin_myo}) and (\ref{eq:source_aniillin_act}), it is tempting to conclude that the net effect of anillin would be to `renormalise' the action of RhoA signalling (Appendix Fig~\ref{fig:RenormalisedKinetics}); leading to effective, $\rho_{\mathrm{A}}$-dependent, concentrations for the effectors mDia1 and ROCK1. Crucially, however, this ignores the fact that anillin has a binding domain for myosin II motors (Fig.~\ref{fig:RhoASignallingPathway}b). Moreover, the binding between anillin and myosin II does not interfere with the action of the key AH domain that scaffolds RhoA signalling. We therefore assume that anillin can be anchored in the cortex by myosin II motors, independently of GTP-RhoA, and, as such, the density of anillin, $\rho_{\mathrm{A}}(x,t)$, is now dictated by
\begin{equation}\label{eq:rhoANLN}
    \partial_t\rho_{\mathrm{A}} + \partial_x (v\rho_{\mathrm{A}}) - D_m\partial_x^2\rho_{\mathrm{A}} = k_{\mathrm{on}}^A(\rho_{\mathrm{m}} - \rho_{\mathrm{A}}) - k_{\mathrm{off}}^A\rho_{\mathrm{A}},
\end{equation}
where anillin can bind at a rate dependent on the number of sites that are available sites on myosin mini-filaments. We impose that actin disassembly cannot exceed the myosin inactivation rate ({\it i.e.}, $k_{\mathrm{off}} > k_{\mathrm{dp}}$) as well as that anillin's unbinding rate is greater than the rate that myosin inactivates ({\it i.e.}, $k_{\mathrm{off}}^A > k_{\mathrm{off}}$).

\begin{table}[t]
\caption{\label{tab:AMAParameters}
Non-dimensionalised parameters used in the linear stability and numerical analysis of the anillin-actomyosin model.}
\begin{ruledtabular}
\begin{tabular}{cc}
Nondimensionalised Parameter &  Rescaled quantity\\
\hline
$\Bar{k}_{\mathrm{off}}^{m}$ & $(k_{\mathrm{on}}(\rho_{\text{ROCK}} + \mu\rho_{\mathrm{A}}^0) +k_{\mathrm{off}}\eta) / \zeta$\\
$\Bar{k}_{\mathrm{on}}^{a}$ &  $k_{\mathrm{on}}\eta\rho_{\mathrm{a}}^0(\rho_{\text{ROCK}} + \mu\rho_{\mathrm{A}}^0) / (\zeta\rho_{\mathrm{m}}^0) $\\
$\Bar{k}_{\mathrm{on}}^{A}$ & $k_{\mathrm{on}}\eta\mu\rho_{\mathrm{A}}^0(\rho_{\mathrm{a}}^0 - \rho_{\mathrm{m}}^0) / (\zeta\rho_{\mathrm{m}}^0)$\\
$\Tilde{k}_{\mathrm{dp}}$& $k_{\mathrm{dp}}\eta / \zeta$\\
$\Hat{k}_{\mathrm{p}}$& $k_{\mathrm{p}}\eta\rho_{\text{mDia}} / (\zeta\rho_{\mathrm{a}}^0)$\\
$\Bar{k}_{\mathrm{p}}$& $k_{\mathrm{p}}\eta\nu\rho_{\mathrm{A}}^0 / (\zeta\rho_{\mathrm{a}}^0)$\\
$\Bar{k}_{\mathrm{off}}^A$& $k_{\mathrm{off}}^A\eta / \zeta$\\
$\Bar{k}_{\mathrm{on}}^A$& $k_{\mathrm{on}}^A\eta\rho_{\mathrm{m}}^0 / (\zeta\rho_{\mathrm{A}}^0)$\\
$\mathrm{Pe}$ & $\zeta / D_m\Gamma$\\
$\Bar{\chi_a}$& $\chi_a\rho_{\mathrm{a}}^0 / \zeta$\\
$h(\rho_{\mathrm{m}}^*)$& $\rho_{\mathrm{m}}^0\rho_{\mathrm{m}}^* / (\rho_{\mathrm{m}}^0+\rho_{\mathrm{m}}^*)^2$
\end{tabular}
\end{ruledtabular}
\end{table}

\subsection{Linear Stability Analysis}

Once again, our first action is to consider the stability of the system about the uniform steady-state solution (given in Appendix~\ref{sec:SuppLinearStability3D}). Non-dimensionalising (see Table \ref{tab:AMAParameters}), we find the following system of linear dynamical equations
\begin{equation}\label{eq:linear_AMA}
\begin{aligned}
  &\partial_t\delta\rho_{\mathrm{m}} + \partial_x\delta v - \mathrm{Pe}^{-1}\partial_x^2\delta\rho_{\mathrm{m}} = \Bar{k}_{\mathrm{on}}^{a}\delta\rho_{\mathrm{a}} + \Bar{k}_{\mathrm{on}}^{A}\delta\rho_{\mathrm{A}} - \Bar{k}_{\mathrm{off}}^{m}\delta\rho_{\mathrm{m}}\textrm{,} \\
  &\partial_t\delta\rho_{\mathrm{a}} + \partial_x \delta v = \Bar{k}_{\mathrm{p}}\delta\rho_{\mathrm{A}} - \Bar{k}_{\mathrm{dp}}\delta\rho_{\mathrm{a}}\textrm{,} \\
  &\partial_t\delta\rho_{\mathrm{A}} + \partial_x \delta v - \mathrm{Pe}^{-1}\partial_x^2\delta\rho_{\mathrm{A}} = \Bar{k}_{\mathrm{o n}}^A\delta\rho_{\mathrm{m}}- \Bar{k}_{\mathrm{off}}^A\delta\rho_{\mathrm{A}}\textrm{,}
\end{aligned}
\end{equation}
which are coupled by the (linearised) relation
\begin{equation}\label{eq:delta_v_AMA}
    \partial_x^2\delta v - \delta v - \Bar{\chi}_a\partial_x\delta\rho_{\mathrm{a}} + h(\rho_{\mathrm{m}}^*) \partial_x \delta\rho_{\mathrm{m}} = 0\textrm{.}
\end{equation}
Here, we have assumed that the rate at which anillin diffuses is identical to that of myosin (diffusion is athermal in our model, being attributed to myosin processivity, and anillin is bound to myosin).

In reciprocal space, (\ref{eq:delta_v_AMA}) can be incorporated into (\ref{eq:linear_AMA}) to result in linear equations of the form,
\begin{equation}
    \partial_t{\bs{\delta\Bar{\rho}}}'_{q} = \mathsf{A}^\prime\cdot{\bs{\delta\Bar{\rho}}}'_{q}\text{,}
\end{equation}
where ${\bs{\delta \Bar{\rho}}}'_{q} = (\delta\Bar{\rho}_{\mathrm{a}}, \delta\Bar{\rho}_{\mathrm{m}}, \delta\Bar{\rho}_{\mathrm{A}})^{\mathrm{T}}$ and 
\begin{widetext}
    \begin{equation}\label{eq:linearResponseMatrix3D}
   \mathsf{A}^\prime =  \left( \begin{matrix}
    -\Tilde{k}_{\mathrm{dp}} -\Tilde{\chi}_a\frac{q^2}{q^2+1} &\frac{q^2 h(\rho_{\mathrm{m}}^*)}{q^2+1} & \Bar{k}_{p}\\
        \Bar{k}_{\mathrm{on}}^{a} -\Tilde{\chi}_a\frac{q^2}{q^2+1} &  -\Bar{k}_{\mathrm{off}}^{m} -\mathrm{Pe}^{-1}  q^2+\frac{q^2 h(\rho_{\mathrm{m}}^*)}{q^2+1} & \Bar{k}_{\mathrm{on}}^{A}\\
         -\Tilde{\chi}_a\frac{q^2}{q^2+1} & q^2+\frac{q^2 h(\rho_{\mathrm{m}}^*)}{q^2+1} + \Bar{k}_{\mathrm{on}}^A & - \Bar{k}_{\mathrm{off}}^A-\mathrm{Pe}^{-1}  q^2
    \end{matrix}\right)
\end{equation}
\end{widetext}

\begin{figure*}[t!]
\centering
\includegraphics[width=\textwidth]{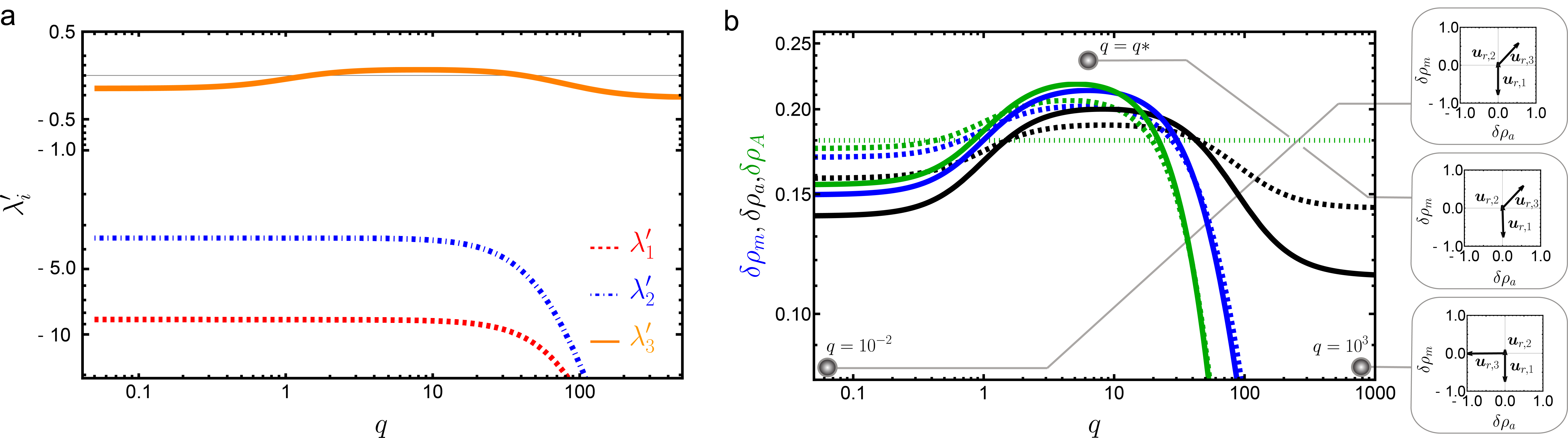}
\caption{\textbf{Linear regime of actomyosin-anillin}. \textbf{a,} Real eigenvalues, $\lambda_{1,2,3}$ of linear response matrix $\mathsf{A'}$ (Eq.~\ref{eq:linearResponseMatrix3D}).
For representative parameter values (Table.~\ref{tab:ParameterValues}) instabilities only occur due to the positivity of the third eigenvalue, which is an modification of the base model's `contractile-kinetic' regime. \textbf{b,} Forward time-evolution of linear response matrix $\mathsf{A}^\prime$ (Eq .~\ref{eq:linearResponseMatrix2D}), plotted for parameters in Table~\ref{tab:ParameterValues}. Dotted, dashed and solid lines denote time points $t = 0$, $t = 1$, and $t = 2$, respectively. Inserts show the 3D projected right-eigenvectors, $\mathbf{u}_r$, plotted for $q = 10^{-2}$ and $q=10^3$ (large and small wavelengths, respectively) as well at the maximum growth rate $q^* \approx 8.0$. Following the baseline model, at wavenumbers both smaller and of similar magnitude to $q^*$, the positivity of $\lambda_3'$ drives changes in both $\delta\rho_{\mathrm{m}}$ and $\delta\rho_{\mathrm{a}}$ (as well as $\delta\rho_{\mathrm{A}}$, see 3D eigenvectors in Fig~\ref{fig:3DFull_Eigenvectors}) which then decouples for larger values of $q$.}
\label{fig:linearResponseAnalysis3D}   
\end{figure*}

Expanding in small $q$ and ignoring terms greater than $\mathcal{O}(q^2)$, we can extract the eigenvalues / vectors and therefore the stability conditions as before. However, the resulting formulas are sufficiently unwieldy that we consign them to Appendix \ref{sec:SuppLinearStability3D}. Nevertheless, by evaluating the expressions for parameter values extracted from the literature (Table \ref{tab:ParameterValues}), we observe qualitatively similar features to our non-anillin model (Fig.~\ref{fig:linearResponseAnalysis3D}). Specifically, we interpret the instability associated with $\lambda_3^\prime$ as a modified contractile-kinetic instability, since it tends to $\lambda_2$ in the limit where the anillin coupling parameters are taken to zero. This is similarly apparent in the projection of the eigenvectors onto the $\delta\rho_{\mathrm{a}}$-$\delta\rho_{\mathrm{m}}$ plane (Fig.~\ref{fig:linearResponseAnalysis3D}c, full three dimensional eigenvectors in Appendix~\ref{sec:SuppLinearStability3D}). For $q\leq q^\ast$, we once again see that equal growth in both actin and myosin, which then decouples for larger values of $q$. This behaviour aligns with the forward linear time evolution (Fig.~\ref{fig:linearResponseAnalysis3D}b) of the system and reveals that the growth rate of actin, myosin (and anillin) for $q \lesssim 10$ have the same $q$-dependence. 

\begin{figure*}[t]
    \centering
    \includegraphics[width=\textwidth]{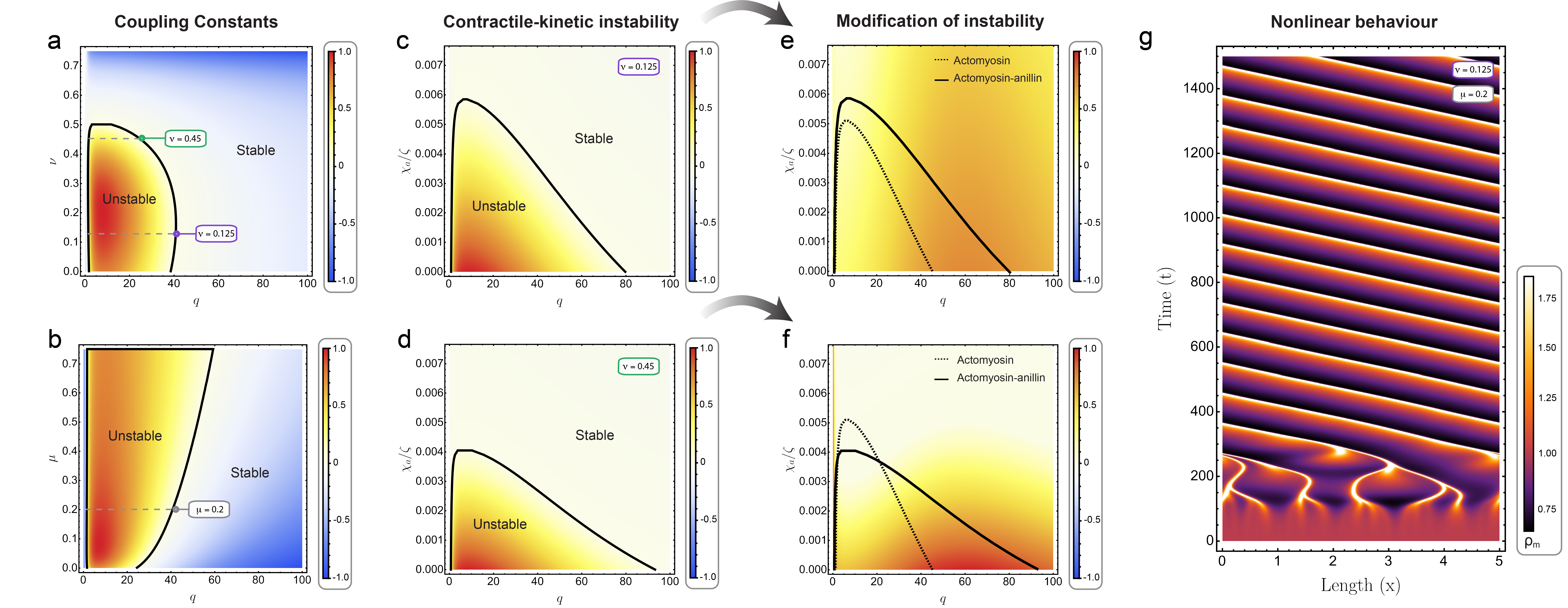}
    \caption{\textbf{Scaffolding actomyosin remodelling}. \textbf{a-b,} Linear stability diagrams showing the anillin dependent linear uplift of (\textbf{a}) actin polymerisation, $\nu$ and (\textbf{b}) myosin activation, $\mu$, as a function of wavenumber, $q$, plotted for representative values of all other parameters (Table ~\ref{tab:ParameterValues}). \textbf{c-d,} For a given $\nu$, the associated linear stability diagrams showing the ratio of inverse compressibility to active contractility, $\chi_a / \zeta$, as a function of wavenumber, $q$, plotted for representative values of all other parameters. \textbf{e-f,} Difference, $\lambda^\prime_3 - \lambda_2$ between the growth rates of the `contractile-kinetic' instabilities associated with actomyosin-anillin and actomyosin-only models, respectively. The unstable regimes of the actomyosin-only and actomyosin-anillin models are denoted by the dashed and dot-dashed black lines, respectively. (\textbf{e}) Actomyosin-anillin anchorage is seen to amplify the existing contractile-kinetic instabilities (illustrated by positive values) as well as inducing instabilities in regimes that would otherwise be stable.  (\textbf{f}) Actomyosin-anillin anchorage is seen to dampen (illustrated by negative values) or amplify existing contractile-kinetic instabilities, as well as inducing instabilities in parameter regimes that would otherwise be stable. \textbf{g,} Kymograph showing the numerical solution of the actomyosin-anillin system. The amplitudes of the initial perturbations of $\rho_{\mathrm{m}}^0$, $\rho_{\mathrm{a}}^0$ and $\rho_{\mathrm{A}}^0$ are drawn uniformly from $A\sim\mathcal{U}[0,0.01)$. The kymograph illustrates an short initial excitable phase followed by a two-peak travelling wave phase.}
    \label{fig:Anillin_dynamics}
\end{figure*}

Despite such similarities, our analysis reveals several key differences that arise due to the coupling to anillin (Fig.~\ref{fig:Anillin_dynamics}). To understand the different ways in which these manifest, it is helpful to focus on the two coupling constants, $\nu$ and $\mu$, which respectively control the linear increase in actin polymerisation and myosin activation with the local density of anillin. Using representative parameter values (Table ~\ref{tab:ParameterValues}) the contractile-kinetic instability can be visualised as a function of $\nu$ and $\mu$ (Fig.~~\ref{fig:Anillin_dynamics}) allowing us to make the following observations.

As $\mu$ increases so does the size of the unstable regime (Fig.~\ref{fig:Anillin_dynamics}b). That is, the greater the (linear) coupling between anillin and myosin, the greater the wavenumber of modes which are unstable.

This is not the case for $\nu$, however, where the size of the unstable regime increases with $\nu$ at low values, but then begins to shrink as $\nu$ continues to increase, until the instability disappears completely (Fig.~\ref{fig:Anillin_dynamics}b).  In other words, as the (linear) coupling between anillin and actin is increased, the maximum wavenumber of the unstable modes first increases, but then decreases until no modes are unstable.

We therefore discern two regimes as to how anillin modifies the nascent contractile-kinetic instability. For low $\nu$ it increases rate at which a particular mode grows, the range of $q$ over which the system is unstable, and the range of $\chi_a/\zeta$ for which the system is unstable, with a lower ratio of contractility to elastic response now being possible (Figs.~\ref{fig:Anillin_dynamics}c \& e).  For higher $\nu$ we still see increases to the rate at which each mode grows, as well as an increased range over which the system in unstable, however the system is only unstable at lower values of $\chi_a/\zeta$, meaning a higher ratio of contractility to elastic response is needed (Figs.~\ref{fig:Anillin_dynamics}d \& f). 

Heuristically, we can understand this behaviour in the the following way. Without anillin, perturbations that increase the local densities of both actin and myosin lead to positive feedback--- {\it i.e.}, a hydrodynamic instability--- if the (dimensionless) ratio of the active contractility to the modulus of inverse compressibility exceeds the (dimensionless) ratio of off- to on-rates of bound myosin (see Section \ref{sec:actomyosin}).  When anillin is present, it is advected with myosin and therefore mirrors any local increases in density.  However, at low values of $\nu/\mu$, this primarily results in an increase in the activation (and therefore on-rate) of myosin, leading to an overall amplification of the instability.  Whereas, at high values of $\nu / \mu$, by contrast, the primary result in an increased polymerisation of actin, which counteracts the effects of the increased density of myosin motors (and therefore requires greater ratio of contractility to inverse compressibility before the system becomes unstable).  


\subsection{Non-linear behaviour}

In order to examine the full non-linear behaviour of our model we once again utilised a central space, RK2-time finite difference scheme. The actin, myosin and anillin densities were initialised with a small perturbation of the form $\delta\rho_{a,m,A} = A\sin\left(2\pi x/L + \theta\right)$ around the homogeneous steady state, where $A\sim \mathcal{U}[0,\text{const.})$ and $\theta\sim \mathcal{U}[0,2\pi)$.

The numerical solution quantitatively aligns with the linear stability analysis over the first $t/\tau<50$ (Appendix Fig.~\ref{fig:3DAlignment}). The solution beyond this point, similar to the actomyosin-only model, forms dense and dynamic foci separated by sparse regions. Again, there is an initial, transient, excitable regime, which is then followed by a stable travelling front regime. In the excitable phase, the newly nucleated foci are attracted to, and merge with larger foci, though these are less prominent (see kymograph in Fig.~\ref{fig:Anillin_dynamics}b and Supplementary  video S2). In addition, the excitable phase is significantly shorter than that of the actomyosin-only model. Analysing the size of the foci reveals that, once fused, the foci return to a characteristic size, before further fusion events occur (Appendix Fig.~\ref{fig:FociWidth}b, left). The foci which lie in the transient excitable regime of the system are notably smaller in width than the actomyosin-only model. This is to be contrasted with the travelling pulse regime, where the foci are larger in both intensity and width, and therefore have a larger characteristic separation (Fig.~\ref{fig:Anillin_dynamics}b, and Appendix Fig.~\ref{fig:FociWidth}b). Lastly, the velocity of the travelling wave is significantly increased in the presence of anillin.

\section{Discussion}

We have used a scalar active hydrodynamic model to capture how actomyosin remodelling is `scaffolded' by the protein anillin, which is found at important sites of cellular contractility, including the cytokinetic furrow and zonula adherens \cite{Budnar2019AnillinKinetics,straight2005anillin,field1995anillin}. By comparing with a base model, where anillin is not present, we are able to highlight how anillin changes the dynamics of remodelling.

At short times, in the linear regime, anillin can significantly alter different aspects of a `contractile-kinetic' instability--- {\it cf.} Eq.~(\ref{eq:eigen2})--- which representative parameter values suggest is the physically relevant phenomenon in a myosin-actin only system. Anillin can alter the range of wavenumbers over which the system is unstable, the values of the key parameter $\tilde{\chi_a}$ ({\it i.e.}, the non-dimensionalised ratio of elastic resistance to compression and active contractility) over which the system is unstable, and the rate of growth of the unstable modes.

At longer times, in the non-linear regime, the faster growth rates induced by anillin lead to a shorter transient phase of excitatory-inhibitory behaviour, forming steady state travelling fronts more quickly than when anillin is absent. Anillin's presence is seen to result in fewer, larger actomyosin foci which travel at faster speeds than the actomyosin-only model.

Here, we remark that, whilst it has already been demonstrated that anillin improves RhoA's efficacy \cite{Budnar2019AnillinKinetics}, this effect alone would, in principle, lead to a `renormalisation' of our anillin-free model (see Appendix Fig.~\ref{fig:RenormalisedKinetics}) via the modification of source sink terms--- \textit{i.e.}, via Eqs.~(\ref{eq:source_anillin_myo}) and (\ref{eq:source_aniillin_act}). It is only when we include anillin's ability to bind to, and hence be advected by, myosin II motors, that the hydrodynamics is materially altered. Our work therefore paves the way for understanding the long wavelength effects of scaffolds and enzymes that, whilst not strictly necessary for regulatory pathways to function, alter the kinetics in such a fashion as to play a meaningful role at a coarse-grained level.

Moreover, in this case, the precise details depend on the values of two coupling parameters, $\nu$ and $\mu$, which quantify the linear relationship(s) between the local density of anillin and the local rates of actin polymerisation and myosin activation, respectively. Whilst the quantity $\mu$ has not been investigated experimentally, we assume that its behaviour may have qualitative similarities with that of $\nu$, which was investigated as part of \cite{Budnar2019AnillinKinetics} using an innovative trans-membrane receptor assay. The implications of that study (see Fig. 5I of \cite{Budnar2019AnillinKinetics}) are that $\nu$ depends on the local concentration of the acidic phospholipid, PIP$_2$, which appears essential for anillin's unique mode of scaffolding. This is a potentially very interesting insight, since it mechanistically relates cortical dynamics and reorganisation to membrane composition.

There are, nevertheless, many ways in which this work might be improved. For example, we do not utilize an RhoA explicit representation, and instead (for simplicity) assume basal rates of RhoA production. Whilst previous work by Budnar \textit{et al.} illustrated that RhoA exchanges rapidly at adherens junctions irrespective of its nucleaotide bound status, other cellular processes such as cytokinesis have noted that specific GEFs/GAPs are critical regulators \cite{chircop2014rho} and such should be taken into account. Further, our model does not incorporate polar ordering of actin filaments, and its effect on myosin-II transport and thus anillin, nor do we consider higher spatial dimensions \cite{staddon2022pulsatile, tan2020topological, tyson1988singular}.

Ultimately, our predictions regarding the actomyosin remodelling and how anillin regulates cortical dynamics open up the opportunity to explore the role of scaffolds and signalling pathways at the long wavelength, hydrodynamic scale. Experimental verification will be key, and we look forward to future work in the area.

\section*{Acknowledgements}
This work is supported by the National Health and Medical Research Council of Australia (1136592 and 1163462), the Australian Research Council (DP220103951) and the European Molecular Biology Organization (EMBO ALTF-251-2018).

\bibliography{bibliography}

\clearpage

\begin{widetext}
\appendix
This appendix consists of three parts (\ref{sec:SuppLinearStability2D}, \ref{sec:SuppLinearStability3D}, and \ref{sec:SuppNumerics}), each divided into multiple subsections. In appendices \ref{sec:SuppLinearStability2D} and \ref{sec:SuppLinearStability3D} we present additional calculational details of the linear stability analysis for the isotropic actomyosin and actomyosin-anillin models. The final appendix (\ref{sec:SuppNumerics}) provides additional information on the numerically solved solutions for the actomyosin and actomyosin-anillin models.

\section{ACTOMYOSIN}\label{sec:SuppLinearStability2D}

\subsubsection*{Eigensystem}
Here, we provide the full-length expressions of the eigensystem (eigenvalues and right-eigenvectors, respectively) of the linear response matrix $\mathsf{A}$ (Eq.~\ref{eq:linearResponseMatrix2D}). The full-length eigenvalues ($\lambda_i$) are given by,
\begin{align}\label{eq:eigenvalues_FL}
    \lambda_{1,2} &= -\frac{1}{2 \left(q^2+1\right)}\Bigg(\Tilde{k}_{\mathrm{off}} +\Tilde{k}_{\mathrm{dp}} +\mathrm{Pe}^{-1}  q^4+\Tilde{k}_{\mathrm{off}}  q^2+\mathrm{Pe}^{-1} q^2+\Tilde{k}_{\mathrm{dp}}  q^2-h(\rho_{\mathrm{m}}^*) q^2+\Tilde{\chi}_a  q^2 \pm \bigg(\Tilde{k}_{\mathrm{dp}} +\mathrm{Pe}^{-1}  q^4\nonumber\\
    &+\Tilde{k}_{\mathrm{off}}  \left(q^2+1\right)+q^2 \left(\mathrm{Pe}^{-1} +\Tilde{k}_{\mathrm{dp}} -h(\rho_{\mathrm{m}}^*) +\Tilde{\chi}_a \right)\bigg)^2-4 \left(q^2+1\right) \bigg(\Tilde{k}_{\mathrm{off}}  \left(\Tilde{k}_{\mathrm{dp}} +\Tilde{k}_{\mathrm{dp}}  q^2+\Tilde{\chi}_a  q^2\right) \nonumber\\
    &+q^2 \left(\mathrm{Pe}^{-1} \left(\Tilde{k}_{\mathrm{dp}} +\Tilde{k}_{\mathrm{dp}}  q^2+\Tilde{\chi}_a  q^2\right)-h(\rho_{\mathrm{m}}^*)  (\Tilde{k}_{\mathrm{dp}} +\Hat{k}_{\mathrm{on}} )\right)\bigg)^{1/2}\Bigg)\text{.}
\end{align}
\\
The associated full-length right-eigenvectors (${\bf u}_{r_i}$) of eigenvalues in Eq.~\ref{eq:eigenvalues_FL} are given by,
\begin{align}\label{eq:eigenvectors_FL}
    {\bf u}_{r_{1,2}} = \left(\frac{\psi \pm \sqrt{\sigma}}{\sqrt{1 + (\psi \pm \sqrt{\sigma})^2}}, \frac{1}{\sqrt{1 + (\psi \pm \sqrt{\sigma})^2}} \right)^{\mathrm{T}}\textrm{,}
\end{align}
where, 
\begin{align}
    \psi =& \frac{1}{2 q^2 \left(\Tilde{\chi}_a -\Hat{k}_{\mathrm{on}} \right)-2 \Hat{k}_{\mathrm{on}}}\left(\Tilde{k}_{\mathrm{dp}} -\Tilde{k}_{\mathrm{off}}  \left(q^2+1\right)+q^2 \left(\Tilde{k}_{\mathrm{dp}} +h(\rho_{\mathrm{m}}^*) +\Tilde{\chi}_a -\mathrm{Pe}^{-1} \left(q^2+1\right)\right)\right)\textrm{,}\nonumber\\
    & \nonumber\\
    \sigma =& \frac{1}{2 q^2 \left(\Tilde{\chi}_a -\Hat{k}_{\mathrm{on}} \right)-2 \Hat{k}_{\mathrm{on}}}\Bigg( (\Tilde{k}_{\mathrm{dp}}) ^2+q^8(\mathrm{Pe}^{-1}) ^2 + 2 q^6  \mathrm{Pe}^{-1}\left(\mathrm{Pe}^{-1} -\Tilde{k}_{\mathrm{dp}} -h(\rho_{\mathrm{m}}^*) -\Tilde{\chi}_a \right) \nonumber\\
    &+q^4 \bigg((\mathrm{Pe}^{-1}) ^2 -2 \mathrm{Pe}^{-1} \bigg(2 \Tilde{k}_{\mathrm{dp}} +h(\rho_{\mathrm{m}}^*) +\Tilde{\chi}_a \bigg)+(\Tilde{k}_{\mathrm{dp}}) ^2+2 \Tilde{k}_{\mathrm{dp}}  \left(h(\rho_{\mathrm{m}}^*) +\Tilde{\chi}_a \right)+\left(h(\rho_{\mathrm{m}}^*) -\Tilde{\chi}_a \right)^2 \nonumber\\
    &+4 h(\rho_{\mathrm{m}}^*)  \Hat{k}_{\mathrm{on}} \bigg)+\Tilde{k}_{\mathrm{off}} ^2 \left(q^2+1\right)^2 +2 q^2 \bigg(-\mathrm{Pe}^{-1}  \Tilde{k}_{\mathrm{dp}} +\Tilde{k}_{\mathrm{dp}}  \left(\Tilde{k}_{\mathrm{dp}} +h(\rho_{\mathrm{m}}^*) +\Tilde{\chi}_a \right)+2 h(\rho_{\mathrm{m}}^*)  \Hat{k}_{\mathrm{on}} \bigg)\nonumber\\
    &+2 \Tilde{k}_{\mathrm{off}} \left(q^2+1\right)\left(-\Tilde{k}_{\mathrm{dp}} +\mathrm{Pe}^{-1}  q^4+q^2 \left(\mathrm{Pe}^{-1} -\Tilde{k}_{\mathrm{dp}} -h(\rho_{\mathrm{m}}^*) -\Tilde{\chi}_a \right)\right)\Bigg)\nonumber\textrm{.}
\end{align}

\subsubsection*{Unphysical Exceptional Point}
Performing a series expansion of the eigenvalues around low $q$ then equating them, the lower critical wavenumber $q'$, for stable spatio-temporal waves occur,
\begin{equation}\label{eq:UnphysicalEP_lengthscale}
    q' = \frac{\sqrt{\Tilde{k}_{\mathrm{dp}} -\Tilde{k}_{\mathrm{off}} }}{\sqrt{\frac{2 h(\rho_{\mathrm{m}}^*)  \Hat{k}_{\mathrm{on}} }{\Tilde{k}_{\mathrm{off}} -\Tilde{k}_{\mathrm{dp}} }+\mathrm{Pe}^{-1} -\Tilde{\chi}_a -h(\rho_{\mathrm{m}}^*) }}.
\end{equation}

Substituting (\ref{eq:UnphysicalEP_lengthscale}) into low $q$ eigenvalue series expansion leads to the following criterion for an exceptional point,
\begin{equation}
    \Tilde{\chi}_a  = \overline{\Hat{k}_{\mathrm{on}}},
\end{equation}

where,
\begin{equation}\label{eq:UnphysicalEP_condition}
     \overline{\Hat{k}_{\mathrm{on}}} = \Hat{k}_{\mathrm{on}}\left(\frac{h(\rho_{\mathrm{m}}^*) }{\Tilde{k}_{\mathrm{off}} -\Tilde{k}_{\mathrm{dp}} }-\frac{\mathrm{Pe}^{-1}}{\Tilde{k}_{\mathrm{off}} -\Tilde{k}_{\mathrm{dp}} -\Hat{k}_{\mathrm{on}}}\right).
\end{equation}

\noindent Note that the length-scale ($q'$) only holds where the actin disassembly rate exceeds the myosin inactivation rate. However, this violates the condition $k_{\mathrm{off}} + k_{\mathrm{on}} - k_{\mathrm{dp}} > 0$ for a biologically feasible setting and such is excluded.

\subsubsection*{Nondimensionalised Parameters \& Values}
\begin{table}[h!]
\caption{List of non-dimensionalised parameters and values used in both the linear stability and non-linear numerical analysis}
    \begin{ruledtabular}
        \begin{tabular}{ccc}
        
            Nondimensionalised Parameter &  Rescaled quantity & Value\\
            \hline
            $\Tilde{k}_{\mathrm{off}}$& $(k_{\mathrm{off}} + k_{\mathrm{on}})\eta / \zeta$ & 2.8\\
            $\Tilde{k}_{\mathrm{p}}$& $k_{\mathrm{p}}\eta / \zeta\rho_{\mathrm{a}}^0$& 0.15\\
            $\Tilde{k}_{\mathrm{dp}}$& $k_{\mathrm{dp}}\eta / \zeta$& 0.15\\
            $\Hat{k}_{\mathrm{on}}$& $k_{\mathrm{on}}\eta\rho_{\mathrm{a}}^0 / \zeta(\rho_{\mathrm{m}}^0)$& 2.8\\
            $\Tilde{D}_m$& $D_m\Gamma / \zeta$& 0.001\\
            $\Tilde{\chi}_a$ & $\chi_a\rho_{\mathrm{a}}^0 / \zeta$& 0.07\\
            $h(\rho_{\mathrm{m}}^*)$& $\rho_{\mathrm{m}}^0\rho_{\mathrm{m}}^* / (\rho_{\mathrm{m}}^0+\rho_{\mathrm{m}}^*)^2$& 0.25\\
        \end{tabular}
    \end{ruledtabular}
\end{table}

\clearpage

\section{ACTOMYOSIN-ANILLIN COUPLING}\label{sec:SuppLinearStability3D}

\subsubsection*{Homogeneous steady state}
In this section we compute the homogeneous steady state solution ($\rho_m^0$, $\rho_a^0$, $\rho_A^0$) of our model. Solving equations ~\ref{eq:source_anillin_myo},\ref{eq:source_aniillin_act} and \ref{eq:rhoANLN} for its steady state solution we obtain the following three simultaneous equations,
\begin{align}
    \rho_{\mathrm{m}}^0 &= \frac{k_{\mathrm{on}}}{k_{\mathrm{off}}}\left(\rho_a^0-\rho_{\mathrm{m}}^0\right)\left(\rho_{\mathrm{ROCK}}+\mu\rho_A^0\right), \label{eq:HSS_rhom}\\
    \rho_{\mathrm{A}}^0 &= \frac{k_{\mathrm{on}}^A}{k_{\mathrm{off}}^A}\rho_{\mathrm{m}}^0, \label{eq:HSS_rhoANLN}\\
    \rho_{\mathrm{a}}^0 &= \frac{k_{\mathrm{p}}}{k_{\mathrm{dp}}}\left(\rho_{\mathrm{mDia}} + \nu\rho_{\mathrm{A}}^0 \right).\label{eq:HSS_rhoa}
\end{align}

Substituting \ref{eq:HSS_rhoa} and \ref{eq:HSS_rhoANLN} into \ref{eq:HSS_rhom} we extract the following single-variable equation,
\begin{equation}
    \frac{k_{\mathrm{on}}}{k_{\mathrm{off}}}\left(\frac{k_{\mathrm{p}}}{k_{\mathrm{dp}}}\left(\rho_{\mathrm{mDia}}+\nu\frac{k_{\mathrm{on}}^A}{k_{\mathrm{off}}^A}\rho_{\mathrm{m}}^0\right)-\rho_{\mathrm{m}}^0\right)\left(\rho_{\mathrm{ROCK}}+\mu\frac{k_{\mathrm{on}}^A}{k_{\mathrm{off}}^A}\rho_{\mathrm{m}}^0\right)-\rho_{\mathrm{m}}^0 = 0.
\end{equation}

This is solved for $\rho_m^ 0$ using Solve in Mathematica with the condition $\rho_m^0 >0$ to yield biologically feasible solutions. The solution for $\rho_m^0$ is then directly substituted into \ref{eq:HSS_rhoa} and \ref{eq:HSS_rhoANLN} for $\rho_a^0$ and $\rho_A^0$, respectively.

\subsubsection*{Eigensystem}
Due to the complexity/length of the eigensystem of the linear response matrix $\mathsf{A'}$ (Eq.~\ref{eq:linearResponseMatrix3D}) we have excluded the full-length expressions for the eigenvalues and right-eigenvectors, respectively. Nevertheless, we are able to compute the stability condition for our system to be unstable for small wavenumbers (\textit{i.e.,} long wavelengths) by retaining terms up to $\mathcal{O}(q^2)$. Here, the stability conditions are respectively given by, 

\begin{align}
    &\psi \left( \frac{2 \sqrt[3]{2}}{\sqrt[3]{\upsilon }} \left(\Bar{k}_{\mathrm{off}}^m +\Bar{k}_{\mathrm{dp}} +\Bar{k}_{\mathrm{off}}^A\right) -1\right) + \frac{\sigma}{\sqrt[3]{2} \upsilon^{2/3}}  - \frac{2\sigma}{2^{2/3} \upsilon ^{4/3}} \omega + \frac{1}{\sqrt[3]{\upsilon }} \phi > 0, \\
    &\psi \left( \frac{\pm 2 \sqrt[3]{\pm2}}{\sqrt[3]{\upsilon }} \left(\Bar{k}_{\mathrm{off}}^m +\Bar{k}_{\mathrm{dp}} +\Bar{k}_{\mathrm{off}}^A\right) -1\right) + \frac{\sigma}{\sqrt[3]{2} \upsilon^{2/3}}  - \frac{2\sigma}{2^{2/3} \upsilon ^{4/3}}(1 \mp i\sqrt{3}) \omega + \frac{1}{\sqrt[3]{\upsilon }}(1\mp i\sqrt{3}) \phi > 0,
\end{align}
\\

\noindent where, $\psi$, $\omega$, $\phi$, $\upsilon$ and $\sigma$ are given by,

\begin{align*}
\psi =& \ \Tilde{k}_{\mathrm{off}} +\Tilde{k}_{\mathrm{dp}} +\Tilde{k}_{\mathrm{off}}^A +\Bar{\chi}_a+2 \ \mathrm{Pe^{-1}}-h(\rho_{\mathrm{m}}^*) \text{,} \\
&\\
\omega =& \ (\Bar{k}_{\mathrm{off}}^m) ^2+(\Bar{k}_{\mathrm{dp}}) ^2+(\Bar{k}_{\mathrm{off}}^A)^2-\Bar{k}_{\mathrm{off}}^m  \left(\Bar{k}_{\mathrm{dp}} +\Bar{k}_{\mathrm{off}}^A \right)-\Bar{k}_{\mathrm{dp}}  \Bar{k}_{\mathrm{off}}^A +3 \ \Bar{k}_{\mathrm{on}}^A  \Bar{k}_{\mathrm{on}}^{A'}\text{,} \\
&\\
\phi =& \ \Bar{k}_{\mathrm{off}}^m \left(\mathrm{Pe}^{-1} +2 \ \Bar{k}_{\mathrm{dp}} +\Bar{\chi}_a +2 \ \Bar{k}_{\mathrm{off}}^A \right)+\mathrm{Pe}^{-1}  \left(2 \ \Bar{k}_{\mathrm{dp}} +\Bar{k}_{\mathrm{off}}^A \right)+\Bar{k}_{\mathrm{off}}^A  \left(2 \ \Bar{k}_{\mathrm{dp}} -h(\rho_{\mathrm{m}}^*) +\Bar{\chi}_a \right)-h(\rho_{\mathrm{m}}^*)  \big(\Bar{k}_{\mathrm{dp}} +\Bar{k}_{\mathrm{on}}^a \\
&+\Bar{k}_{\mathrm{on}}^A\big)+\Bar{\chi}_a \Bar{k}_{\mathrm{p}}\text{,}\\
&\nonumber\\
\upsilon =& \ -2 \left(\Bar{k}_{\mathrm{off}}^m\right) ^3+3 \left(\Bar{k}_{\mathrm{off}}^m\right) ^2 \left(\Bar{k}_{\mathrm{dp}} +\Bar{k}_{\mathrm{off}}^A \right)+3 \ \Bar{k}_{\mathrm{off}}^m  \left(\left(\Bar{k}_{\mathrm{dp}}\right) ^2-4 \ \Bar{k}_{\mathrm{dp}}  \Bar{k}_{\mathrm{off}}^A +\left(\Bar{k}_{\mathrm{off}}^A\right) ^2-3 \ \Bar{k}_{\mathrm{on}}^{A'}  \Bar{k}_{\mathrm{on}}^A \right)-2 \left(\Bar{k}_{\mathrm{dp}}\right) ^3 \\
&+3 \left(\Bar{k}_{\mathrm{dp}}\right) ^2 \Bar{k}_{\mathrm{off}}^A+3 \ \Bar{k}_{\mathrm{dp}}  \left(\left(\Bar{k}_{\mathrm{off}}^A\right) ^2+6 \ \Bar{k}_{\mathrm{on}}^{A'}  \Bar{k}_{\mathrm{on}}^A \right)+27 \ \Bar{k}_{\mathrm{on}}^a  \Bar{k}_{\mathrm{p}}  \Bar{k}_{\mathrm{on}}^A -2 \left(\Bar{k}_{\mathrm{off}}^A\right) ^3-9 \  \Bar{k}_{\mathrm{off}}^A  \Bar{k}_{\mathrm{on}}^{A'}  \Bar{k}_{\mathrm{on}}^A + 3 \Bigg(-3 \left(\Bar{k}_{\mathrm{on}}^A\right) ^2 \\
& \bigg(\left(\Bar{k}_{\mathrm{off}}^m\right) ^2 \left(\Bar{k}_{\mathrm{on}}^{A'}\right) ^2+2 \left(\Bar{k}_{\mathrm{off}}^m\right)  \left(\Bar{k}_{\mathrm{on}}^{A'}\right)\bigg(4 \ \Bar{k}_{\mathrm{dp}}  \left(\Bar{k}_{\mathrm{on}}^{A'}\right) +9 \ \Bar{k}_{\mathrm{on}}^a  \Bar{k}_{\mathrm{p}} -5 \ \Bar{k}_{\mathrm{off}}^A  \Bar{k}_{\mathrm{on}}^{A'} \bigg)-8 \left(\Bar{k}_{\mathrm{dp}}\right) ^2 \left(\Bar{k}_{\mathrm{on}}^{A'}\right) ^2+2 \  \Bar{k}_{\mathrm{off}}^A  \Bar{k}_{\mathrm{on}}^{A'}  \\
&\left(4 \Bar{k}_{\mathrm{dp}}  \Bar{k}_{\mathrm{on}}^{A'} +9 \Bar{k}_{\mathrm{on}}^a  \Bar{k}_{\mathrm{p}} \right)-36 \ \Bar{k}_{\mathrm{dp}}  \Bar{k}_{\mathrm{on}}^a  \Bar{k}_{\mathrm{p}}  \Bar{k}_{\mathrm{on}}^{A'} -27 \left(\Bar{k}_{\mathrm{on}}^a\right) ^2 \left(\Bar{k}_{\mathrm{p}}\right) ^2+\left(\Bar{k}_{\mathrm{off}}^A\right) ^2 \left(\Bar{k}_{\mathrm{on}}^{A'}\right) ^2\bigg) -6 \ \Bar{k}_{\mathrm{on}}^A  \Bigg(\left(\Bar{k}_{\mathrm{off}}^m\right) ^3 \Bar{k}_{\mathrm{dp}}  \Bar{k}_{\mathrm{on}}^{A'} \\
& +2 \left(\Bar{k}_{\mathrm{off}}^m\right) ^3 \Bar{k}_{\mathrm{on}}^a  \Bar{k}_{\mathrm{p}}+\left(\Bar{k}_{\mathrm{off}}^A\right) ^2 \bigg(\left(\Bar{k}_{\mathrm{on}}^{A'}\right)  \left(4 \left(\Bar{k}_{\mathrm{off}}^m\right) ^2-5 \ \Bar{k}_{\mathrm{off}}^m  \Bar{k}_{\mathrm{dp}} +\left(\Bar{k}_{\mathrm{dp}}\right) ^2\right)-3 \Bar{k}_{\mathrm{on}}^a  \Bar{k}_{\mathrm{p}}  \left(\Bar{k}_{\mathrm{off}}^m +\Bar{k}_{\mathrm{dp}} \right)\bigg)- \Bar{k}_{\mathrm{off}}^A\bigg(3 \ \Bar{k}_{\mathrm{on}}^a  \Bar{k}_{\mathrm{p}} \\
&\left(\left(\Bar{k}_{\mathrm{off}}^m\right) ^2-4 \ \Bar{k}_{\mathrm{off}}^m  \Bar{k}_{\mathrm{dp}} +\left(\Bar{k}_{\mathrm{dp}}\right) ^2\right)+\Bar{k}_{\mathrm{on}}^{A'}  \left(\Bar{k}_{\mathrm{off}}^m -\Bar{k}_{\mathrm{dp}} \right) \big(\left(\Bar{k}_{\mathrm{off}}^m\right) ^2+6 \ \Bar{k}_{\mathrm{off}}^m  \Bar{k}_{\mathrm{dp}} -4 \left(\Bar{k}_{\mathrm{dp}}\right) ^2\big)\bigg)+\left(\Bar{k}_{\mathrm{off}}^m\right) ^2 \left(\Bar{k}_{\mathrm{dp}}\right) ^2 \Bar{k}_{\mathrm{on}}^{A'}\\
&-3 \left(\Bar{k}_{\mathrm{off}}^m\right) ^2 \Bar{k}_{\mathrm{dp}}  \Bar{k}_{\mathrm{on}}^a  \Bar{k}_{\mathrm{p}} -4 \ \Bar{k}_{\mathrm{off}}^m  \left(\Bar{k}_{\mathrm{dp}}\right) ^3 \Bar{k}_{\mathrm{on}}^{A'} -3 \ \Bar{k}_{\mathrm{off}}^m  \left(\Bar{k}_{\mathrm{dp}}\right) ^2 \Bar{k}_{\mathrm{on}}^a  \Bar{k}_{\mathrm{p}}+\left(\Bar{k}_{\mathrm{off}}^A\right) ^3 \left(-\Bar{k}_{\mathrm{off}}^m  \Bar{k}_{\mathrm{on}}^{A'}+\Bar{k}_{\mathrm{dp}}  \Bar{k}_{\mathrm{on}}^{A'}+2 \ \Bar{k}_{\mathrm{on}}^a  \Bar{k}_{\mathrm{p}} \right)\\
&+2 \left(\Bar{k}_{\mathrm{dp}}\right) ^4 \Bar{k}_{\mathrm{on}}^{A'} +2 \left(\Bar{k}_{\mathrm{dp}}\right) ^3 \Bar{k}_{\mathrm{on}}^a  \Bar{k}_{\mathrm{p}} \Bigg) -3 \left(\Bar{k}_{\mathrm{off}}^m -\Bar{k}_{\mathrm{dp}} \right)^2 \left(\Bar{k}_{\mathrm{off}}^m -\Bar{k}_{\mathrm{off}}^A \right)^2 \left(\Bar{k}_{\mathrm{dp}} -\Bar{k}_{\mathrm{off}}^A \right)^2-12 \left(\Bar{k}_{\mathrm{on}}^{A'}\right) ^3 \left(\Bar{k}_{\mathrm{on}}^A\right) ^3\Bigg)^{1/2},\\
&\\
\sigma &= -2 \Bigg(\Bar{k}_{\mathrm{off}}^m \bigg(2 \ \mathrm{Pe}^{-1} \left(\Bar{k}_{\mathrm{dp}}-2 \ \Bar{k}_{\mathrm{off}}^A\right)-3 (\Bar{k}_{\mathrm{dp}})^2+2 \ \Bar{k}_{\mathrm{dp}} \left(h(\rho_{\mathrm{m}}^*) +6 \ \Bar{k}_{\mathrm{off}}^A-\Bar{\chi}_a\right)-3 \left(\Bar{k}_{\mathrm{off}}^A\right)^2+2 \ h(\rho_{\mathrm{m}}^*)  \Bar{k}_{\mathrm{off}}^A\\
&+4 \ \Bar{k}_{\mathrm{off}}^A \Bar{\chi}_a+3 \ h(\rho_{\mathrm{m}}^*)  \Bar{k}_{\mathrm{on}}^a+3 \ h(\rho_{\mathrm{m}}^*)  \Bar{k}_{\mathrm{on}}^A+9 \ \Bar{k}_{\mathrm{on}}^A \Bar{k}_{\mathrm{on}}^{A'}+6 \ \Bar{k}_{\mathrm{p}} \Bar{\chi}_a\bigg)+\mathrm{Pe}^{-1} \bigg(-2 \left(\Bar{k}_{\mathrm{dp}}\right)^2+2 \ \Bar{k}_{\mathrm{dp}} \Bar{k}_{\mathrm{off}}^A+\left(\Bar{k}_{\mathrm{off}}^A\right)^2\\
&+6 \ \Bar{k}_{\mathrm{on}}^A \Bar{k}_{\mathrm{on}}^{A'}\bigg)+\left(\Bar{k}_{\mathrm{off}}^m\right)^2 \left(\mathrm{Pe}^{-1}-2 \ h(\rho_{\mathrm{m}}^*) -3 \ \Bar{k}_{\mathrm{dp}}-3 \ \Bar{k}_{\mathrm{off}}^A-\Bar{\chi}_a\right)+2 \left(\Bar{k}_{\mathrm{dp}}\right)^3+h(\rho_{\mathrm{m}}^*)  \left(\Bar{k}_{\mathrm{dp}}\right)^2-3 \left(\Bar{k}_{\mathrm{dp}}\right)^2 \Bar{k}_{\mathrm{off}}^A\\
&+2 \ (\Bar{k}_{\mathrm{dp}})^2 \Bar{\chi}_a-3 \ \Bar{k}_{\mathrm{dp}} \left(\Bar{k}_{\mathrm{off}}^A\right)^2-4 \ h(\rho_{\mathrm{m}}^*)  \Bar{k}_{\mathrm{dp}} \Bar{k}_{\mathrm{off}}^A-3 \ \Bar{k}_{\mathrm{on}}^{A'} \bigg(6 \ \Bar{k}_{\mathrm{dp}} \Bar{k}_{\mathrm{on}}^A+\Bar{k}_{\mathrm{on}}^A \left(h(\rho_{\mathrm{m}}^*) -3 \ \Bar{k}_{\mathrm{off}}^A+2 \ \Bar{\chi}_a\right)\\
&-3 \ \Bar{k}_{\mathrm{p}} \left(\Bar{\chi}_a-3 \ \Bar{k}_{\mathrm{on}}^a\right)\bigg)-2 \ \Bar{k}_{\mathrm{dp}} \Bar{k}_{\mathrm{off}}^A \Bar{\chi}_a+3 \ h(\rho_{\mathrm{m}}^*)  \Bar{k}_{\mathrm{dp}} \Bar{k}_{\mathrm{on}}^a-6 \ h(\rho_{\mathrm{m}}^*)  \Bar{k}_{\mathrm{dp}} \Bar{k}_{\mathrm{on}}^A-3 \ \Bar{k}_{\mathrm{dp}} \Bar{k}_{\mathrm{p}} \Bar{\chi}_a+2 \left(\Bar{k}_{\mathrm{off}}^A\right)^3\\
&+h(\rho_{\mathrm{m}}^*)  \left(\Bar{k}_{\mathrm{off}}^A\right)^2-\left(\Bar{k}_{\mathrm{off}}^A\right)^2 \Bar{\chi}_a-6 \ h(\rho_{\mathrm{m}}^*)  \Bar{k}_{\mathrm{off}}^A \Bar{k}_{\mathrm{on}}^a+3 \ h(\rho_{\mathrm{m}}^*)  \Bar{k}_{\mathrm{off}}^A \Bar{k}_{\mathrm{on}}^A-3 \ \Bar{k}_{\mathrm{off}}^A \Bar{k}_{\mathrm{p}} \Bar{\chi}_a+2 \left(\Bar{k}_{\mathrm{off}}^m\right)^3-9 \ h(\rho_{\mathrm{m}}^*)  \Bar{k}_{\mathrm{on}}^a \Bar{k}_{\mathrm{p}}\Bigg)\\
&+\Bigg(-3 \left(\Bar{k}_{\mathrm{on}}^{A'}\right)^2 \bigg(-8 (\Bar{k}_{\mathrm{dp}})^2 \left( \Bar{k}_{\mathrm{on}}^A\right)^2+2 \  \Bar{k}_{\mathrm{off}}^m  \Bar{k}_{\mathrm{on}}^A \left(4 \ \Bar{k}_{\mathrm{dp}}  \Bar{k}_{\mathrm{on}}^A-5 \  \Bar{k}_{\mathrm{off}}^A  \Bar{k}_{\mathrm{on}}^A+9 \  \Bar{k}_{\mathrm{on}}^a \Bar{k}_{\mathrm{p}}\right)+4 \ \Bar{k}_{\mathrm{dp}}  \Bar{k}_{\mathrm{on}}^A \big(2 \  \Bar{k}_{\mathrm{off}}^A  \Bar{k}_{\mathrm{on}}^A\\
&-9 \ \Bar{k}_{\mathrm{on}}^a \Bar{k}_{\mathrm{p}}\big)+\left( \Bar{k}_{\mathrm{off}}^A\right)^2 \left( \Bar{k}_{\mathrm{on}}^A\right)^2+18 \  \Bar{k}_{\mathrm{off}}^A  \Bar{k}_{\mathrm{on}}^a \Bar{k}_{\mathrm{on}}^A  \Bar{k}_{\mathrm{p}}+\left( \Bar{k}_{\mathrm{off}}^m\right)^2 \left( \Bar{k}_{\mathrm{on}}^A\right)^2-27 \left(\Bar{k}_{\mathrm{on}}^a\right)^2  \Bar{k}_{\mathrm{p}}^2\bigg)-6 \ \Bar{k}_{\mathrm{on}}^{A'} \bigg(2 \left(\Bar{k}_{\mathrm{dp}}\right)^4  \Bar{k}_{\mathrm{on}}^A\\
&-4 \left(\Bar{k}_{\mathrm{dp}}\right)^3  \Bar{k}_{\mathrm{off}}^m  \Bar{k}_{\mathrm{on}}^A+2 \left(\Bar{k}_{\mathrm{dp}}\right)^3  \Bar{k}_{\mathrm{on}}^a \Bar{k}_{\mathrm{p}}+\left( \Bar{k}_{\mathrm{off}}^A\right)^2 \bigg( \Bar{k}_{\mathrm{on}}^A \bigg(\left(\Bar{k}_{\mathrm{dp}}\right)^2-5 \ \Bar{k}_{\mathrm{dp}}  \Bar{k}_{\mathrm{off}}^m+4 \left( \Bar{k}_{\mathrm{off}}^m\right)^2\bigg)-3 \  \Bar{k}_{\mathrm{on}}^a \Bar{k}_{\mathrm{p}} \big(\Bar{k}_{\mathrm{dp}} \\
&+ \Bar{k}_{\mathrm{off}}^m\big)\bigg)-\Bar{k}_{\mathrm{off}}^A \bigg(3 \  \Bar{k}_{\mathrm{on}}^a \Bar{k}_{\mathrm{p}} \left(\left(\Bar{k}_{\mathrm{dp}}\right)^2-4 \ \Bar{k}_{\mathrm{dp}}  \Bar{k}_{\mathrm{off}}^m+\left( \Bar{k}_{\mathrm{off}}^m\right)^2\right)+ \Bar{k}_{\mathrm{on}}^A \left( \Bar{k}_{\mathrm{off}}^m-\Bar{k}_{\mathrm{dp}}\right)\bigg(-4 \left(\Bar{k}_{\mathrm{dp}}\right)^2+6 \ \Bar{k}_{\mathrm{dp}}  \Bar{k}_{\mathrm{off}}^m\\
&+\left( \Bar{k}_{\mathrm{off}}^m\right)^2\bigg)\bigg)+\left(\Bar{k}_{\mathrm{dp}}\right)^2 \left( \Bar{k}_{\mathrm{off}}^m\right)^2  \Bar{k}_{\mathrm{on}}^A-3 \left(\Bar{k}_{\mathrm{dp}}\right)^2  \Bar{k}_{\mathrm{off}}^m  \Bar{k}_{\mathrm{on}}^a \Bar{k}_{\mathrm{p}}+\left( \Bar{k}_{\mathrm{off}}^A\right)^3 \left(\Bar{k}_{\mathrm{dp}}  \Bar{k}_{\mathrm{on}}^A- \Bar{k}_{\mathrm{off}}^m  \Bar{k}_{\mathrm{on}}^A+2 \  \Bar{k}_{\mathrm{on}}^a \Bar{k}_{\mathrm{p}}\right)\\
&+\Bar{k}_{\mathrm{dp}} \left( \Bar{k}_{\mathrm{off}}^m\right)^3  \Bar{k}_{\mathrm{on}}^A-3 \ \Bar{k}_{\mathrm{dp}} \left( \Bar{k}_{\mathrm{off}}^m\right)^2  \Bar{k}_{\mathrm{on}}^a \Bar{k}_{\mathrm{p}}+2 \left( \Bar{k}_{\mathrm{off}}^m\right)^3  \Bar{k}_{\mathrm{on}}^a \Bar{k}_{\mathrm{p}}\bigg)-3 \left(\Bar{k}_{\mathrm{dp}}- \Bar{k}_{\mathrm{off}}^A\right)^2 \left( \Bar{k}_{\mathrm{off}}^m-\Bar{k}_{\mathrm{dp}}\right)^2 \left( \Bar{k}_{\mathrm{off}}^m- \Bar{k}_{\mathrm{off}}^A\right)^2\\
&-12 \left( \Bar{k}_{\mathrm{on}}^A\right)^3 \left(\Bar{k}_{\mathrm{on}}^{A'}\right)^3\Bigg)^{1/2}\Bigg(12\bigg(\left(\Bar{k}_{\mathrm{dp}}+ \Bar{k}_{\mathrm{off}}^A+ \Bar{k}_{\mathrm{off}}^m\right)^2-3 \ \Bar{k}_{\mathrm{off}}^m \left(\Bar{k}_{\mathrm{dp}}+ \Bar{k}_{\mathrm{off}}^A\right)-3 \ \Bar{k}_{\mathrm{dp}}  \Bar{k}_{\mathrm{off}}^A+3 \ \Bar{k}_{\mathrm{on}}^A \Bar{k}_{\mathrm{on}}^{A'}\bigg)^2 \\
&\bigg(3 \bigg( \Bar{k}_{\mathrm{off}}^m \left(\mathrm{Pe}^{-1}+2 \ \Bar{k}_{\mathrm{dp}}+2 \  \Bar{k}_{\mathrm{off}}^A+\Bar{\chi}_a\right)+\mathrm{Pe}^{-1} \left(2 \ \Bar{k}_{\mathrm{dp}}+ \Bar{k}_{\mathrm{off}}^A\right)-h(\rho_{\mathrm{m}}^*) \Bar{k}_{\mathrm{dp}}-h(\rho_{\mathrm{m}}^*)  \Bar{k}_{\mathrm{off}}^A-h(\rho_{\mathrm{m}}^*) \Bar{k}_{\mathrm{on}}^a\\
&-h(\rho_{\mathrm{m}}^*)  \Bar{k}_{\mathrm{on}}^A+2 \ \Bar{k}_{\mathrm{dp}}  \Bar{k}_{\mathrm{off}}^A+ \Bar{k}_{\mathrm{off}}^A \Bar{\chi}_a-2 \ \Bar{k}_{\mathrm{on}}^A \Bar{k}_{\mathrm{on}}^{A'}+ \Bar{k}_{\mathrm{p}} \Bar{\chi}_a\bigg)-2 \left(\Bar{k}_{\mathrm{dp}}+ \Bar{k}_{\mathrm{off}}^A+ \Bar{k}_{\mathrm{off}}^m\right) \big(2 \ \mathrm{Pe}^{-1}-h(\rho_{\mathrm{m}}^*)+\Bar{k}_{\mathrm{dp}}+ \Bar{k}_{\mathrm{off}}^A\\
&+ \Bar{k}_{\mathrm{off}}^m+\Bar{\chi}_a\big)\bigg)-6 \bigg(-2 (\Bar{k}_{\mathrm{dp}})^3+3 \ \Bar{k}_{\mathrm{off}}^m \bigg(\left(\Bar{k}_{\mathrm{dp}}\right)^2-4 \ \Bar{k}_{\mathrm{dp}}  \Bar{k}_{\mathrm{off}}^A+\left( \Bar{k}_{\mathrm{off}}^A\right)^2-3 \  \Bar{k}_{\mathrm{on}}^A \Bar{k}_{\mathrm{on}}^{A'}\bigg)+3 \left(\Bar{k}_{\mathrm{dp}}\right)^2  \Bar{k}_{\mathrm{off}}^A+3 \ \Bar{k}_{\mathrm{dp}} \\
&\left(\left( \Bar{k}_{\mathrm{off}}^A\right)^2+6 \ \Bar{k}_{\mathrm{on}}^A \Bar{k}_{\mathrm{on}}^{A'}\right)+3 \left( \Bar{k}_{\mathrm{off}}^m\right)^2 \left(\Bar{k}_{\mathrm{dp}}+ \Bar{k}_{\mathrm{off}}^A\right)-2 \left( \Bar{k}_{\mathrm{off}}^A\right)^3-9 \ \Bar{k}_{\mathrm{off}}^A  \Bar{k}_{\mathrm{on}}^A \Bar{k}_{\mathrm{on}}^{A'}-2 \left( \Bar{k}_{\mathrm{off}}^m\right)^3+27 \  \Bar{k}_{\mathrm{on}}^a\Bar{k}_{\mathrm{on}}^{A'}  \Bar{k}_{\mathrm{p}}\bigg) \bigg( \Bar{k}_{\mathrm{off}}^m \\
&\bigg(2 \ \mathrm{Pe}^{-1} \left(\Bar{k}_{\mathrm{dp}}-2 \  \Bar{k}_{\mathrm{off}}^A\right)+2 \ \Bar{k}_{\mathrm{dp}} \big(h(\rho_{\mathrm{m}}^*)+6 \  \Bar{k}_{\mathrm{off}}^A-\Bar{\chi}_a\big)+2 \ h(\rho_{\mathrm{m}}^*)  \Bar{k}_{\mathrm{off}}^A+3 \ h(\rho_{\mathrm{m}}^*) \Bar{k}_{\mathrm{on}}^a+3 \ h(\rho_{\mathrm{m}}^*)  \Bar{k}_{\mathrm{on}}^A-3 \left(\Bar{k}_{\mathrm{dp}}\right)^2\\
&-3 \left( \Bar{k}_{\mathrm{off}}^A\right)^2+4 \ \Bar{k}_{\mathrm{off}}^A \Bar{\chi}_a+9 \ \Bar{k}_{\mathrm{on}}^A \Bar{k}_{\mathrm{on}}^{A'}+6 \ \Bar{k}_{\mathrm{p}} \Bar{\chi}_a\bigg)+\left( \Bar{k}_{\mathrm{off}}^m\right)^2 \left(\mathrm{Pe}^{-1}-2 \ h(\rho_{\mathrm{m}}^*)-3 \ \Bar{k}_{\mathrm{dp}}-3 \ \Bar{k}_{\mathrm{off}}^A-\Bar{\chi}_a\right)\\
&+\mathrm{Pe}^{-1} \bigg(-2 \left(\mathrm{Pe}^{-1}\Bar{k}_{\mathrm{dp}}\right)^2+2 \ \Bar{k}_{\mathrm{dp}}  \Bar{k}_{\mathrm{off}}^A+\left( \Bar{k}_{\mathrm{off}}^A\right)^2+6 \ \Bar{k}_{\mathrm{on}}^A \Bar{k}_{\mathrm{on}}^{A'}\bigg)+h(\rho_{\mathrm{m}}^*) \left(\Bar{k}_{\mathrm{dp}}\right)^2-3 \ \Bar{k}_{\mathrm{on}}^{A'} \bigg( \Bar{k}_{\mathrm{on}}^A \left(h(\rho_{\mathrm{m}}^*)-3  \Bar{k}_{\mathrm{off}}^A+2 \Bar{\chi}_a\right)\\
&+6 \ \Bar{k}_{\mathrm{dp}}  \Bar{k}_{\mathrm{on}}^A-3 \ \Bar{k}_{\mathrm{p}} \left(\Bar{\chi}_a-3 \ \Bar{k}_{\mathrm{on}}^a\right)\bigg)-4 \ h(\rho_{\mathrm{m}}^*) \Bar{k}_{\mathrm{dp}}  \Bar{k}_{\mathrm{off}}^A+3 \ h(\rho_{\mathrm{m}}^*) \Bar{k}_{\mathrm{dp}} \Bar{k}_{\mathrm{on}}^a-6 \ h(\rho_{\mathrm{m}}^*) \Bar{k}_{\mathrm{dp}}  \Bar{k}_{\mathrm{on}}^A+h(\rho_{\mathrm{m}}^*) \left( \Bar{k}_{\mathrm{off}}^A\right)^2\\
&-6 \ h(\rho_{\mathrm{m}}^*)  \Bar{k}_{\mathrm{off}}^A \Bar{k}_{\mathrm{on}}^a+3 \ h(\rho_{\mathrm{m}}^*)  \Bar{k}_{\mathrm{off}}^A  \Bar{k}_{\mathrm{on}}^A-9 \ h(\rho_{\mathrm{m}}^*) \Bar{k}_{\mathrm{on}}^a \Bar{k}_{\mathrm{p}}+2 \left(\Bar{k}_{\mathrm{dp}}\right)^3-3 (\Bar{k}_{\mathrm{dp}})^2  \Bar{k}_{\mathrm{off}}^A+2 \left(\Bar{k}_{\mathrm{dp}}\right)^2 \Bar{\chi}_a-3 \ \Bar{k}_{\mathrm{dp}} \left( \Bar{k}_{\mathrm{off}}^A\right)^2\\
&-2 \ \Bar{k}_{\mathrm{dp}}  \Bar{k}_{\mathrm{off}}^A \Bar{\chi}_a-3 \ \Bar{k}_{\mathrm{dp}}  \Bar{k}_{\mathrm{p}} \Bar{\chi}_a+2 \left( \Bar{k}_{\mathrm{off}}^A\right)^3-\left( \Bar{k}_{\mathrm{off}}^A\right)^2 \Bar{\chi}_a-3 \Bar{k}_{\mathrm{off}}^{A}  \Bar{k}_{\mathrm{p}} \Bar{\chi}_a+2 \left( \Bar{k}_{\mathrm{off}}^m\right)^3\bigg)\Bigg).
\end{align*}

\subsubsection*{Nondimensionalised Parameters \& Values}
\begin{table}[h!]
\caption{List of non-dimensionalised parameters and values used in both the linear stability and non-linear numerical analysis}
    \begin{ruledtabular}
        \begin{tabular}{ccc}
            Nondimensionalised Parameter &  Rescaled quantity & Value\\
            \hline
            $\Bar{k}_{\mathrm{off}}^{m}$ & $(k_{\mathrm{on}}(\rho_{\text{ROCK}} + \mu\rho_{\mathrm{A}}^0) +k_{\mathrm{off}}\eta) / \zeta$ & 8.26\\
            $\Bar{k}_{\mathrm{on}}^{a}$ & $k_{\mathrm{on}}\eta\rho_{\mathrm{a}}^0(\rho_{\text{ROCK}} + \mu\rho_{\mathrm{A}}^0) / (\zeta\rho_{\mathrm{m}}^0) $ & 8.26\\
            $\Bar{k}_{\mathrm{on}}^{A}$ & $k_{\mathrm{on}}\eta\mu\rho_{\mathrm{A}}^0(\rho_{\mathrm{a}}^0 - \rho_{\mathrm{m}}^0) / (\zeta\rho_{\mathrm{m}}^0)$ & 0.43\\
            $\Tilde{k}_{\mathrm{dp}}$& $k_{\mathrm{dp}}\eta / \zeta$ & 0.15\\
            $\Hat{k}_{\mathrm{p}}$& $k_{\mathrm{p}}\eta\rho_{\text{mDia}} / (\zeta\rho_{\mathrm{a}}^0)$ & 0.13\\
            $\Bar{k}_{\mathrm{p}}$& $k_{\mathrm{p}}\eta\nu\rho_{\mathrm{A}}^0 / (\zeta\rho_{\mathrm{a}}^0)$ & 0.02\\
            $\Bar{k}_{\mathrm{off}}^A$& $k_{\mathrm{off}}^A\eta / \zeta$ & 3.8\\
            $\Bar{k}_{\mathrm{on}}^A$& $k_{\mathrm{on}}^A\eta\rho_{\mathrm{m}}^0 / (\zeta\rho_{\mathrm{A}}^0)$ & 3.8\\
            $1 / \mathrm{Pe}$& $D_m\Gamma/\zeta$ & 0.001\\
            $\Bar{\chi_a}$& $\chi_a\rho_{\mathrm{a}}^0/\zeta$ & 0.08\\
            $h(\rho_{\mathrm{m}}^*)$& $\rho_{\mathrm{m}}^0\rho_{\mathrm{m}}^* / (\rho_{\mathrm{m}}^0+\rho_{\mathrm{m}}^*)^2$ & 0.25\\
        \end{tabular}
    \end{ruledtabular}    
\end{table}

\clearpage

\subsubsection*{Three-dimensional right-eigenvectors}
Our response matrix $\mathsf{A'}$ is non-Hermitian and such the right eigenvectors $\boldsymbol{u}_{r,1}$, $\boldsymbol{u}_{r,2}$ and $\boldsymbol{u}_{r,3}$ to which the stability conditions pertain are non-orthogonal. Following the actomyosin model, the difference between the directions of the eigenvectors depends on $q$. Taking representative values from the literature (Table ~\ref{tab:ParameterValues}), we plot the right-eigenvectors for both small and large $q$ as well as the fastest growing mode $q^*$.

\begin{figure*}[h!]
\centering
\includegraphics[width=\textwidth]{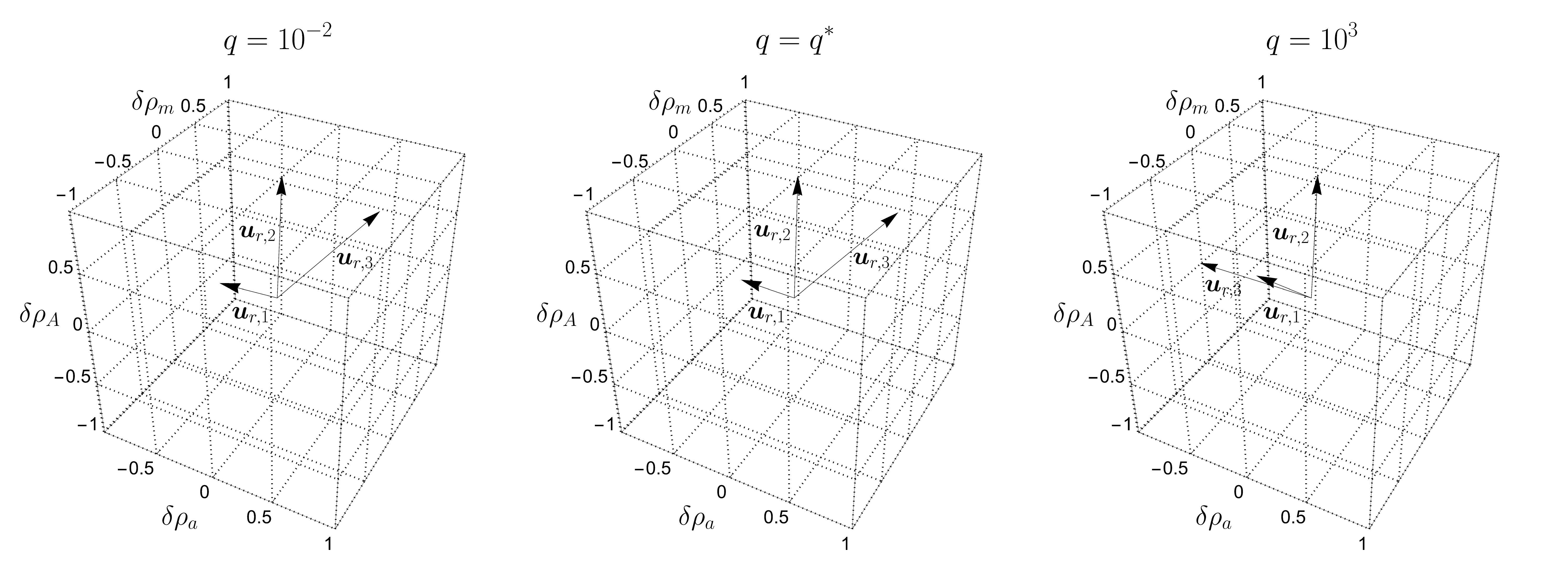}\label{fig:Supp3DEigenvectors}
\caption{Right-eigenvectors, $\mathbf{u}_r$, plotted for parameters in Table~\ref{tab:ParameterValues} at $q = 10^{-2}$ and $q=10^3$ (small and large wavelengths, respectively) as well as the maximum growth rate $q^\ast$. For wavenumbers of both smaller and of similar magnitude to $q^\ast$, the positivity of $\lambda_3'$ drives changes for $\delta\rho_{\mathrm{a}}$, $\delta\rho_{\mathrm{m}}$ and $\delta\rho_{\mathrm{A}}$ (\textit{i.e.,} $\mathbf{u}_3$ forms an angle of approximately $\pi/2$ with the positive x- and z- axis).}
\label{fig:3DFull_Eigenvectors}   
\end{figure*}

For wavenumbers of both smaller and of similar magnitude to $q^\ast$, eigenvector $\mathbf{u}_3$ which is associated with the instability ($\lambda_3'$) forms an angle of approximately $\pi/2$ with the positive x- and z- axis and such when unstable, perturbations to $\delta\rho_{\mathrm{a}}$, $\delta\rho_{\mathrm{m}}$ and $\delta\rho_{\mathrm{A}}$ will grow at approximately similar rates. At large wavenumbers (\textit{i.e.,} short wavelengths) actin and myosin decouple.

\subsubsection*{Renormalised Kinetics}
Here, we analyse the effects of renormalised kinetics of our anillin-free model via the modification of the source and sink terms (via Eqs.~\ref{eq:source_anillin_myo} and ~\ref{eq:source_aniillin_act}) where $\rho_A$ is now a constant value. Taking representative values for the literature (Table ~\ref{tab:ParameterValues}), we plot linear stability diagrams to illustrate the effects of renormalised kinetics for variable $\mu$, $\nu$ and constant $\rho_A$.
\\

\begin{figure*}[h!]
\centering
\includegraphics[width=\textwidth]{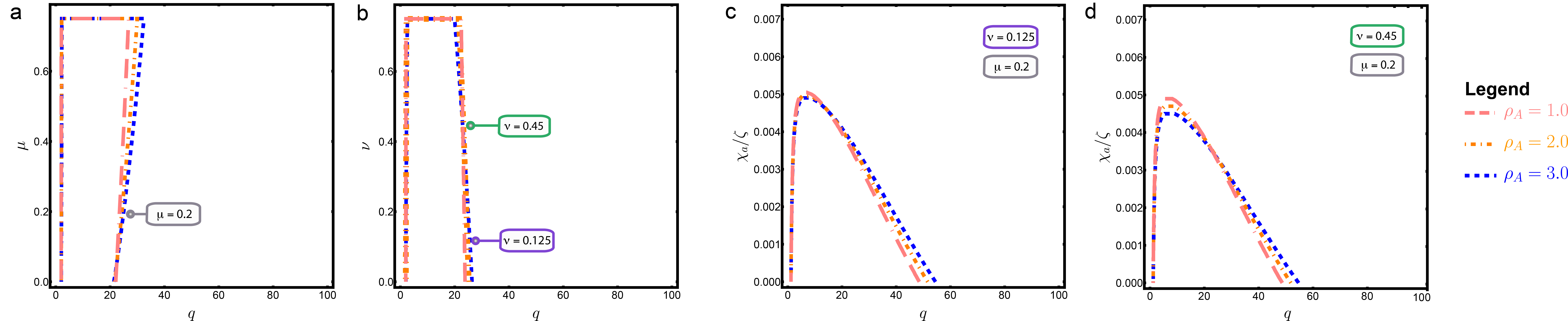}
\caption{\textbf{Renormalised Kinetics.} \textbf{a-b,} Linear stability diagrams illustrating the anillin dependent linear uplift of (\textbf{a}) actin polymerisation, $\nu$) and (\textbf{b}) myosin activation, $\mu$, as a function of wavenumber, $q$, plotted for representative values of all other parameters (Table XX). \textbf{c-d,} For a given $\nu$ and $\mu$, the associated linear stability diagrams, plotted for representative values in Table~\ref{tab:ParameterValues}, showing the ratio of inverse compressibility to active contractility, $\chi_a/\zeta$, as a function of wavenumber, $q$. Dashed, dot-dashed and dot-dot-dashed lines denote anillin values $\rho_A = 3.0$, $\rho_A = 2.0$, $\rho_A = 1.0$, respectively. }
\label{fig:RenormalisedKinetics}   
\end{figure*}

Examining the effect of renormalised kinetics revealed that increases in $\mu$ (Fig.~\ref{fig:RenormalisedKinetics}b) increases the size of the unstable regime. This behaviour is amplified as anillin density increases. Conversely, increasing $\nu$ (Fig.~\ref{fig:RenormalisedKinetics}a) results in an dampening of the unstable regime, however, will never completely stabilise the system. Increases in anillin density acts to either amplify (for $\nu \ll$) or diminish (for $\nu \gg$) the size of unstable regime. Lastly, increasing anillin density, regardless of chosen $\mu$ and $\nu$, acts to dampen previous instabilities as well as inducing instabilities in previously stable $q$-space (Fig.~\ref{fig:RenormalisedKinetics}c,d).

\clearpage

\section{NUMERICS}\label{sec:SuppNumerics}

\subsubsection*{Alignment}
To show the numerical solution of the growth of the initial perturbations quantitatively aligns to the linear stability analysis, we solved our models for 500 (actomyosin) and 350 (actomyosin-anillin) simulations, respectively. Following, we applied the 1-D Discrete Fourier Transform (using python SciPy package FFT \cite{2020SciPy-NMeth}) to the change in density at each time step (\textit{i.e.,} $\delta\rho_i(t_n) = \rho_i(t_n) - \rho_i^0$) which we then average across the total number of simulations to yield the average density growth which we define here as $\delta\rho_i^{\mathrm{nm}}(t_n)$. Figures ~\ref{fig:2DAlignment} and ~\ref{fig:3DAlignment} illustrates several representative time points of the linear stability and numerical solution from the actomyosin and actomyosin-anillin models, respectively.
\\

\textbf{Actomyosin}
\begin{figure*}[h!]
\centering
\includegraphics[width=\textwidth]{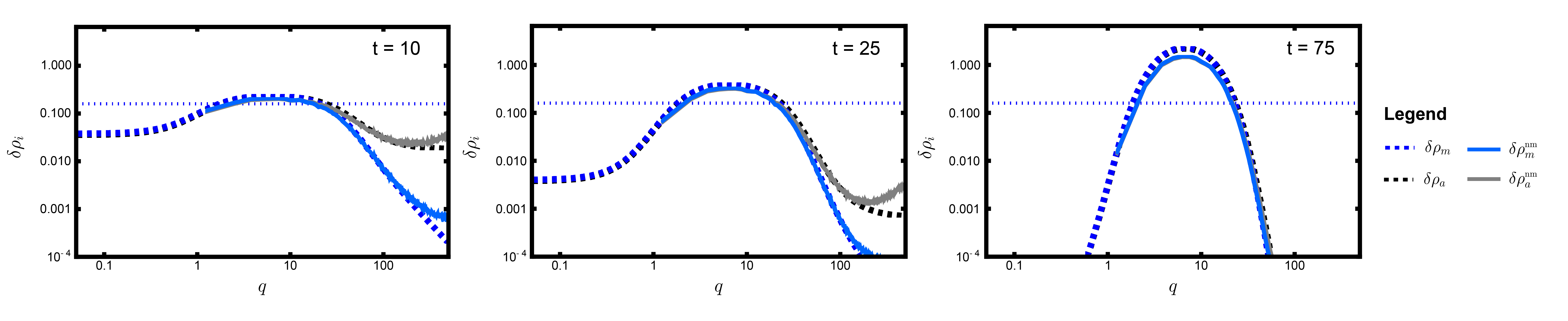}
\caption{Forward time evolution of the linear response matrix, $\mathsf{A}$ [Eq .~(\ref{eq:linearResponseMatrix2D})] and the numerical solution over the first $t/\tau<75$ where $\tau = \zeta/\eta$ is the viscous dissipation associated with contractile remodelling. Panels (left to right) illustrate time points $t= 10, 25$ and $75$. Legend: $\rho_i^{\text{nm}}$ denotes the numerical solution.}
\label{fig:2DAlignment}   
\end{figure*}

\textbf{Actomyosin-anillin anchorage}
\begin{figure*}[h!]
\centering
\includegraphics[width=\textwidth]{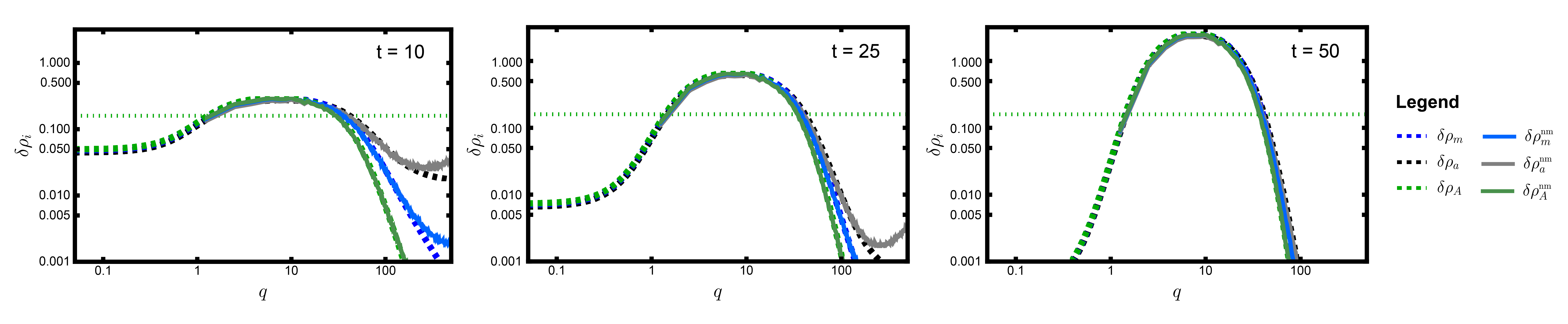}
\caption{Forward time evolution of the linear response matrix, $\mathsf{A'}$ [Eq .~(\ref{eq:linearResponseMatrix3D})] and the numerical solution over the first $t/\tau<50$ where $\tau = \zeta/\eta$ is the viscous dissipation associated with contractile remodelling. Panels (left to right) illustrate time points $t= 10, 25$ and $50$. Legend: $\rho_i^{\text{nm}}$ denotes the numerical solution.}
\label{fig:3DAlignment}   
\end{figure*}

\clearpage
\subsubsection*{Foci Width}
We observe that the foci formed during the transient excitable regime and travelling wave were of variable width in both models. To analyse the foci size we utilise the functions find\_peaks, peak\_widths, peak\_prominences from the python SciPy package Signal \cite{2020SciPy-NMeth}.

\begin{figure*}[h!]
\centering
\includegraphics[width=\textwidth]{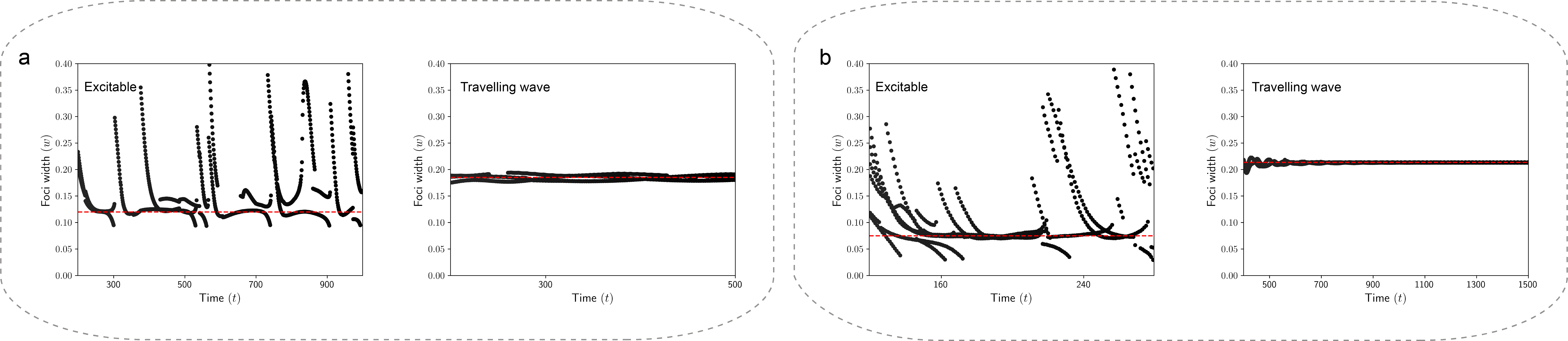}
\caption{Time evolution of the foci width during the transient excitable regime and in the travelling wave front. In the transient excitable regime, foci consistency return to a characteristic width following merging events. Once the travelling wave front is established the foci width increases and stabilises. \textbf{a,} Actomyosin. The foci have a characteristic width of 0.13 and 0.19 in the excitable and travelling wave front regimes (denoted by the red dashed lines), respectively \textbf{b,} Actomyosin-anillin anchorage. The foci have a characteristic width of 0.075 and 0.215 in the excitable and travelling wave front regimes (denoted by the red dashed lines), respectively.}
\label{fig:FociWidth}   
\end{figure*}
\clearpage

\clearpage
\end{widetext}
\end{document}